\documentclass[12pt]{article}
\usepackage[utf8]{inputenc}
\usepackage[margin=1in]{geometry}
\usepackage{natbib}
\usepackage{graphicx}
\usepackage{placeins}
\usepackage{amsmath}
\usepackage{multirow}
\usepackage{xcolor}
\usepackage{setspace}
\usepackage[hidelinks]{hyperref}

\title{Path-specific causal decomposition analysis with multiple correlated mediator variables}
\author{Melissa J. Smith, Leslie A. McClure, D. Leann Long}
\date{}

\begin{document}

\maketitle

\raggedbottom

\section*{Abstract}
A causal decomposition analysis allows researchers to determine whether the difference in a health outcome between two groups can be attributed to a difference in each group's distribution of one or more modifiable mediator variables. With this knowledge, researchers and policymakers can focus on designing interventions that target these mediator variables. Existing methods for causal decomposition analysis either focus on one mediator variable or assume that each mediator variable is conditionally independent given the group label and the mediator-outcome confounders. In this paper, we propose a flexible causal decomposition analysis method that can accommodate multiple correlated and interacting mediator variables, which are frequently seen in studies of health behaviors and studies of environmental pollutants. We extend a Monte Carlo-based causal decomposition analysis method to this setting by using a multivariate mediator model that can accommodate any combination of binary and continuous mediator variables. Furthermore, we state the causal assumptions needed to identify both joint and path-specific decomposition effects through each mediator variable. To illustrate the reduction in bias and confidence interval width of the decomposition effects under our proposed method, we perform a simulation study. We also apply our approach to examine whether differences in smoking status and dietary inflammation score explain any of the Black-White differences in incident diabetes using data from a national cohort study.

\section{Introduction}
Causal decomposition analysis methods can be used to study health disparities and investigate whether different interventions can help reduce these disparities. A causal decomposition analysis allows researchers to determine whether the difference in a health outcome between two groups can be attributed to a difference in each group's distribution of one or more modifiable mediator variables \citep{sudharsanan2021educational, jackson2018decomposition}. With this knowledge, researchers and policymakers can focus on designing interventions that target these mediator variables. One question that may be explored with a causal decomposition analysis is whether matching the mediator distribution of one group to the distribution of the other group can narrow the gap in health outcomes between the two groups. In this paper, we focus on this intervention, which we refer to as ``mediator equalization.''

While a causal decomposition analysis shares similar terminology with a causal mediation analysis, a key distinction is that the group label is held fixed when computing decomposition contrasts based on counterfactual outcomes \citep{sudharsanan2021educational}. Thus, decomposition effects are not synonymous to the total, direct, and indirect effects typically calculated in a mediation analysis. By not changing the group label to compute causal decomposition effects, researchers do not need to contend with the debated topic of ``no manipulation, no causation.'' This debate centers around whether counterfactual outcomes can be defined based on a change in an intrinsic or non-manipulable characteristic such as race or rural-urban status \citep{vanderweele2014causal, glymour2017evaluating}. 

Secondly, the intervention to equalize the mediator distributions between the two populations or groups represents a population-level intervention rather than an individual-level intervention \citep{moreno2018understanding}. This intervention circumvents the need to define counterfactual mediators for each individual based on non-manipulable group labels. This idea of population-level interventions is at the core of so-called interventional mediation effects \citep{moreno2018understanding}, which were extended to the multiple mediator setting by \cite{vansteelandt2017interventional}. A decomposition analysis differs, though, in that only confounders of the mediator-outcome relationship need to be included in the analysis \citep{sudharsanan2021educational}. This relaxes some of the strong no unmeasured confounding assumptions required to perform a causal mediation analysis.

\cite{jackson2018decomposition} introduced the concept of a causal decomposition analysis, and \cite{sudharsanan2021educational} proposed a flexible causal decomposition method using the g-computation technique from causal inference. Furthermore, \cite{smith2023} introduced a Bayesian method for performing causal decomposition analyses with spatially-correlated disease counts. In the work by \cite{sudharsanan2021educational}, the authors focused on an instructive example involving a single mediator variable. However, \cite{sudharsanan2020rural} employed this method using multiple mediators. They examined whether rural-urban differences in adult life expectancy in Indonesia could be attributed to differences in blood pressure, BMI, and smoking. In their study, rural residents experienced lower life expectancy compared to urban residents, so their goal was to determine whether interventions on blood pressure, BMI, and smoking,  individually and jointly, could reduce this difference in life expectancy. 

It is important to highlight that to model the mediator variables, \cite{sudharsanan2020rural} fit three separate regression models, implicitly assuming that the mediators were independent of one another given rural-urban status and the confounders of the mediator-outcome relationships.  In addition to presenting a joint effect of simultaneously equalizing all of the mediators in the rural population to be the same as those in the urban population, they presented path-specific effects, which may exhibit bias in the presence of residual correlation. The authors also did not include mediator-mediator interactions in their outcome model. While it is unclear if residual correlation or interaction effects were present in this study, correlated and potentially interacting mediator variables frequently occur in many types of studies. For example, studies that involve examining the mediating effect of multiple health behaviors have the potential for residual correlation \citep{schuit2002clustering, laaksonen2002associations} as do studies examining the mediating effect of multiple environmental exposures (e.g. \cite{kim2019bayesian}, \cite{bellavia2019approaches}, \cite{roberts2022persistent}). Note that in our paper, the term ``residual correlation'' represents the remaining dependence of the mediators beyond what can be explained by the covariates and group label in the mediator model(s). For two continuous mediators modeled with separate linear regression models, this directly corresponds to the correlation in the models' residuals.

In this paper, we propose a flexible causal decomposition analysis method that accounts for the residual correlation and interaction effects found in many real datasets. Our method extends the decomposition analysis method proposed by \cite{sudharsanan2021educational} to accommodate more complex settings in which there are multiple, correlated, contemporaneous mediator variables. We describe our method, the causal assumptions needed to identify both path-specific and joint decomposition effects, and our method's implementation in R. We also thoroughly compare our proposed method and the existing method in a simulation study to illustrate settings where each method is advantageous.

The remainder of this paper is organized as follows. In Section \ref{section:existing_method}, we introduce relevant notation along with the existing method proposed by \cite{sudharsanan2021educational}. In Section \ref{section:methods}, we describe our extension of this method to accommodate the setting of multiple correlated mediator variables measured contemporaneously. We present a simulation study to compare the multiple mediator implementation of the method by \cite{sudharsanan2021educational} found in \cite{sudharsanan2020rural} to our proposed method in Section \ref{section:simulation} under different data-generating scenarios. We then illustrate our method's application by examining racial differences in diabetes incidence in the REasons for Geographic and Racial Differences in Stroke (REGARDS) Study in Section \ref{section:application}. In Section \ref{section:discussion}, we conclude with a discussion of the strengths, limitations, and further applications of our method.

\section{Existing causal decomposition analysis method}
\label{section:existing_method}
\begin{figure}[ht]
\begin{center}
\includegraphics[scale = 1.1]{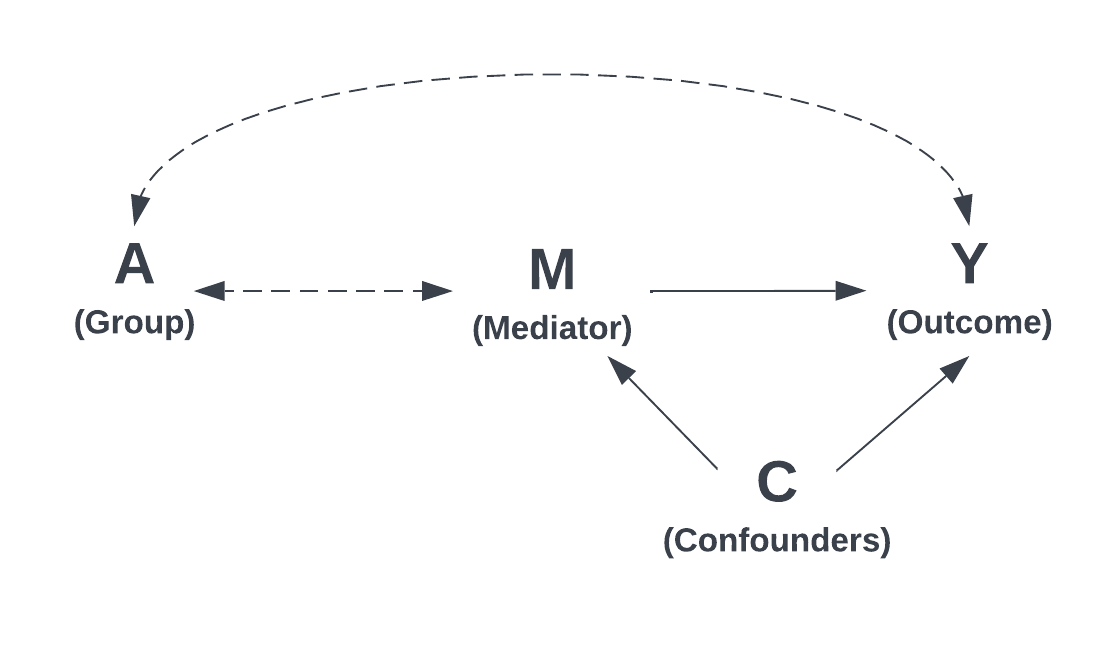}
\end{center}
\caption{Graph depicting the causal and associative relationships between variables in a causal decomposition analysis adapted from the paper by \cite{sudharsanan2021educational}. The solid arrows indicate hypothesized causal relationships whereas the dotted bidirectional arrows indicate associative, but not necessarily causal relationships.}
\label{fig:decomp_existing}
\end{figure}

In the paper by \cite{sudharsanan2021educational}, the authors display two graphs illustrating the similarities and differences between a causal decomposition analysis and a causal mediation analysis. Their graph depicting variable relationships in a causal decomposition analysis is much like Figure \ref{fig:decomp_existing}. This figure depicts the assumed relationships between variables in the decomposition analysis, where $Y$ denotes an individual's observed health outcome, $A$ denotes the random variable corresponding to the individual's group label, $M$ denotes the individual's observed mediator value, and $\boldsymbol{C}$ is a vector that contains any mediator-outcome confounders. For simplicity, we will assume that $A$ is binary and coded as either a 0 or 1, with $A=1$ corresponding to the group with the higher disease burden and $A=0$ corresponding to the group with the lower disease burden. Note that $A$ does not need to be binary and could include more than two groups in practice.

Like the figure by \cite{sudharsanan2021educational}, the bidirectional dotted arrows between the group and outcome, as well as between the group and mediator, indicate associations between these variables. These associations could either represent presumed causal relationships or could be due to non-causal backdoor paths, for example, historical or geographical inequities. On the other hand, the solid directed arrows represent assumed causal relationships. The only causal relationships assumed in the causal decomposition analysis method proposed by \cite{sudharsanan2021educational} are the relationships between $M$, $Y$, and the effects of $C$ on $M$ and $Y$. If the relationship between $A$ and $Y$ is thought to be causal, but the research team does not want to define counterfactual outcomes based on changes in an intrinsic characteristic, causal decomposition analysis methods can still be employed. \cite{jackson2018decomposition} used solid directed arrows between race and an outcome, yet they did not attempt to estimate causal effects for this relationship.

The causal decomposition analysis method proposed by \cite{sudharsanan2021educational} uses a Monte Carlo simulation-based approach for computing the decomposition effects of interest. Their method involves generating various hypothetical populations under the natural mediator values and under mediator values corresponding to the ``mediator equalization'' intervention. Using the outcomes generated for each pseudopopulation, a summary of the health outcome difference between groups can be computed. These summary measures can then be compared in the natural-course and counterfactual pseudopopulations to determine whether the intervention has an effect in reducing the gap in health outcomes between groups. Broadly, the steps of their decomposition analysis method are described below:
\begin{enumerate}
    \item Fit an outcome model that relates the outcome to the group label, mediator, and mediator-outcome confounders.
    \item Fit a model relating the mediator value to the group label and mediator-outcome confounders.
    \item Generate a natural-course pseudopopulation. This pseudopopulation is created by plugging in each individual's observed group labels and confounders into the mediator model and obtaining random draws of the mediator variable for each individual. Then, the random mediator draws are plugged into the outcome model, and random draws are simulated for each individual's ``natural-course'' outcome.
    \item Compute a measure of the disparity of the outcome between the individuals in the two groups of interest using the natural-course outcomes. This could be a relative risk, for example, between the two groups of interest.
    \item Generate counterfactual pseudopopulations corresponding to each intervention of interest. Under the mediator equalization intervention, the mediator values for group $A=1$ may be equalized to the mediator values in group $A=0$. Then, these new mediator values would be plugged into the outcome model, and counterfactual outcomes would be generated. 
    \item Compute the selected disparity measure between individuals in the two groups of interest using each of the counterfactual pseudopopulations. 
    \item Summarize how the disparity measure changes between the natural-course and counterfactual pseudopopulations as a measure of the intervention effect.
    \item Repeat the simulation-based steps a large number of times to reduce the Monte Carlo error in the effect estimates.
\end{enumerate}
This method is highly flexible when a single mediator variable is of interest. It allows for any type of outcome and mediator variable to be used in the decomposition analysis. It also allows for many different types of health summary measures to be specified. As mentioned previously, the paper by \cite{sudharsanan2020rural} implemented this method to assess the effect of three mediator variables in explaining the rural-urban difference in life expectancy in Indonesia. They modified the outcome model in Step 1 to include effects for all three mediators and Step 2 to fit three separate mediator models. They then generated numerous counterfactual pseudopopulations corresponding to an intervention that equalized all of the mediators in the rural population jointly and interventions that equalized one or two mediators at a time to assess path-specific effects. While this method is quite flexible in the absence of residual correlation between mediators, the path-specific effects may be biased in this setting.  Our extended methodology seeks to accurately identify path-specific decomposition effects in the presence of correlated and possibly interacting mediator variables. 

\section{Proposed method for causal decomposition analysis with correlated and interacting mediator variables} \label{section:methods}

\begin{figure}[ht]
\begin{center}
\includegraphics[scale = 1.1]{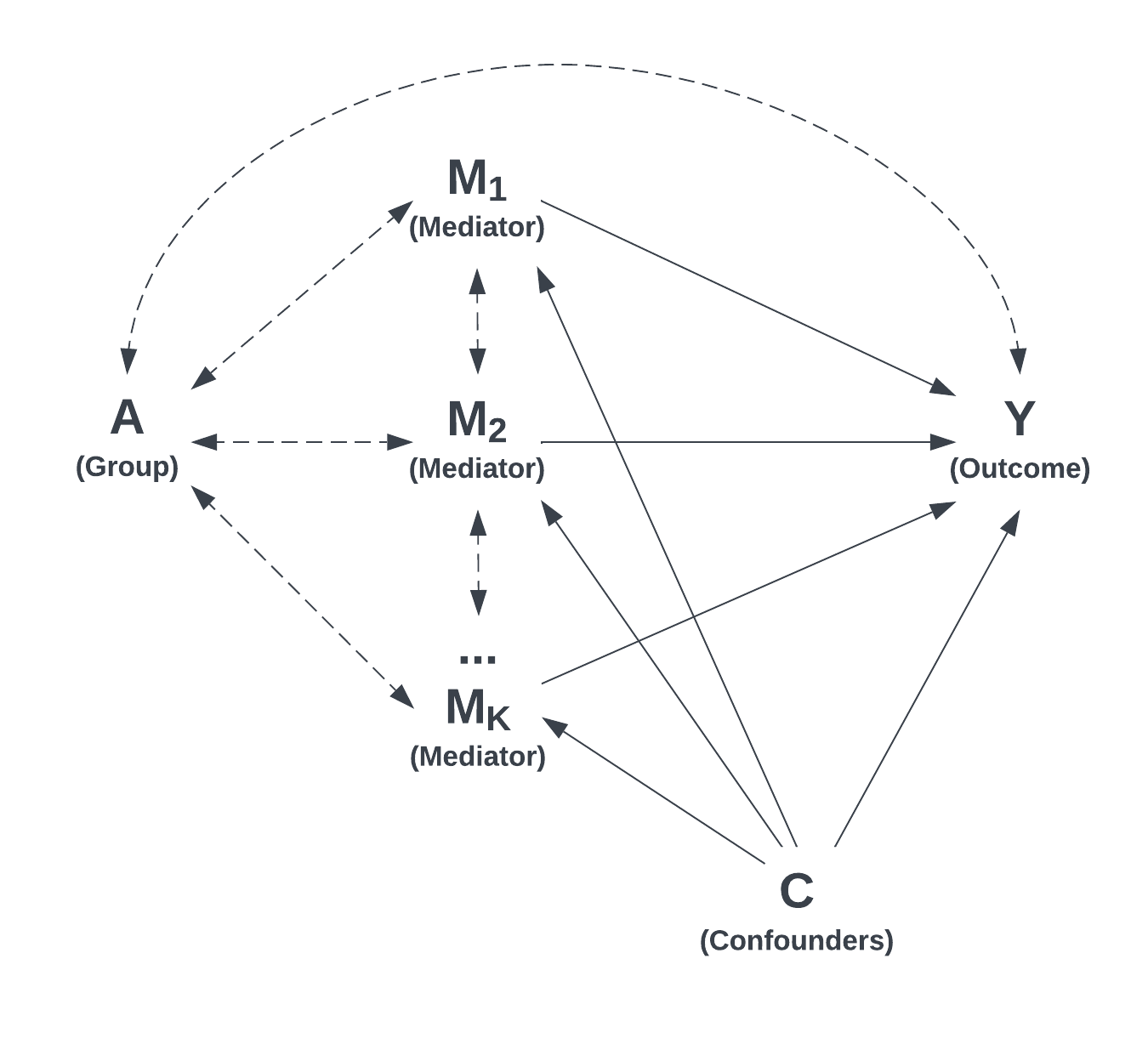}
\end{center}
\caption{Graph depicting the causal and associative relationships between variables in a causal decomposition analysis with multiple correlated mediator variables. The solid arrows indicate hypothesized causal relationships whereas the dotted bidirectional arrows indicate associative, but not necessarily causal relationships.}
\label{fig:decomp_multiple_mediators}
\end{figure}

Figure \ref{fig:decomp_multiple_mediators} displays the setting described in which the mediators are assumed to be correlated with one another and contemporaneous in nature. Let $\boldsymbol{M} = [M_{1},M_{2}, \dots M_K]'$ be a vector of an individual's $K$ observed correlated mediator values. This correlation may be induced by an underlying unmeasured cause, $U$. Similar to Figure \ref{fig:decomp_existing}, $A$ is assumed to be associated with both $\boldsymbol{M}$ and $Y$, but the nature of these associations is left unspecified. However, one important restriction on the associative relationships between $A$, $\boldsymbol{M}$, and $Y$ is that $A$ must not be a collider that opens a backdoor path between $Y$ and $\boldsymbol{M}$. When $A$ is suspected to be a collider variable, the method applied in \cite{sudharsanan2020rural} and our extension should not be used, as the joint and path-specific effects will be subject to bias.

For simplicity, we will assume that an individual's observed health outcome, $Y$, is binary such that $Y \in \{0,1\}.$ Without loss of generalizability, we will focus on the setting of two correlated mediator variables, so $\boldsymbol{M} = [M_{1},M_{2}]'$ represents the vector of mediators for each individual, with $M_1$ corresponding to the first mediator and $M_2$ corresponding to the second mediator. Each mediator may be binary or continuous. Note that our method naturally extends to more than two correlated mediators, and these mediators may be any combination of binary and continuous variables.

\subsection{Counterfactual notation}
 We will next define the counterfactual and conditional notation used to perform our decomposition analysis. Let parentheses denote counterfactual outcomes and $f_{(\cdot\mid\cdot)}$ denote a conditional probability distribution. $Y(A, \boldsymbol{m})$ is the counterfactual outcome when $A$ is held at its natural value and $\boldsymbol{M}$ is set to $\boldsymbol{m}$. For each individual, we always leave the group label fixed at its observed value.

Rather than specifying counterfactual mediators, we instead equalize the mediator distributions between groups $A = 0$ and $A = 1$ within strata of $\boldsymbol{C}$. To account for correlation between the mediators, we assume that they arise from a joint distribution. Thus, the notation $\boldsymbol{M} \sim f_{\boldsymbol{M}\mid A_1 = a_1, A_2 = a_2, \boldsymbol{C} = \boldsymbol{c}} = f_{\boldsymbol{M}\mid a_1, a_2, \boldsymbol{C}}$ indicates that $\boldsymbol{M}$ follows a joint distribution where the distribution of $M_1$ is equalized to that of group $a_1 \in \{0,1\}$ and the distribution of $M_2$ is simultaneously equalized to that of group $a_2 \in \{0,1\}$ within the strata of observed covariates, $\boldsymbol{C}$. Combining the counterfactual and conditional notation, $Y(A, \boldsymbol{M} \sim f_{\boldsymbol{M}\mid a_1, a_2, \boldsymbol{C}})$ is the counterfactual outcome when the group label is held fixed at the observed level, but the mediator values are randomly drawn from the joint distribution which equalizes the first mediator to group $a_1$'s distribution and the second mediator to group $a_2$'s distribution.

For individuals in group $A = 1$, $Y(1, \boldsymbol{M} \sim f_{\boldsymbol{M}\mid 1, 1, \boldsymbol{C}})$ represents the natural-course outcome, since $A$, $a_1$, and $a_2$ are all equal to the natural group label. In contrast, $Y(1, \boldsymbol{M} \sim f_{\boldsymbol{M}\mid 0, 0, \boldsymbol{C}})$, $Y(1, \boldsymbol{M} \sim f_{\boldsymbol{M}\mid 0, 1, \boldsymbol{C}})$, and $Y(1, \boldsymbol{M} \sim f_{\boldsymbol{M}\mid 1, 0, \boldsymbol{C}})$ represent counterfactual outcomes corresponding to different interventions. The first intervention equalizes $M_1$ and $M_2$ in group $A = 1$ to be similar to the distributions in group $A = 0$ and can be conceptualized as a joint intervention. The second and third counterfactual outcomes equalize just one of the two mediators, while accounting for the correlation between $M_1$ and $M_2$ and can be conceptualized as counterfactual outcomes corresponding to path-specific interventions.

\subsection{Outcome model}
To model our binary outcome, $Y$, we use a logistic regression model. Assuming there exists a mediator-mediator interaction, our outcome model can be specified as:
\begin{align*}
Y &\sim \text{Bernoulli}(P(Y = 1)) \\
\text{logit}(P(Y = 1)) &= \beta_0 + \beta_1A + \beta_2M_1 + \beta_3M_2 + \beta_4M_1M_2 + \boldsymbol{\beta}_5'\boldsymbol{C}
\end{align*}
Note that this is one of many ways the model can be specified. Our approach is highly flexible, and any model that relates $Y$ to $A$, $\boldsymbol{M}$, and $\boldsymbol{C}$ may be used as the outcome model. For example, mediator-confounder interactions may be included, as done by \cite{sudharsanan2021educational}.

\subsection{Joint mediator model} \label{sec:joint}
Next, we propose modelling the mediators jointly using a multivariate mediator model rather than two univariate models. In this joint model, we regress $\boldsymbol{M}$ on $A$ and $\boldsymbol{C}$, so that the mediator distributions of one group can be equalized to that of the other group within strata of $\boldsymbol{C}$. We also account for the unobserved confounder or source of residual correlation between the mediators in order to identify the path-specific decomposition effects. 

To model the joint distribution of the two mediators, we apply the joint mediator models described in \cite{wang2013estimation} and \cite{nguyen2016causal} to this decomposition analysis. Assuming both mediators are continuous, we can specify a bivariate linear regression model that relates $M_1$ and $M_2$ to $A$ and $\boldsymbol{C}$ while accounting for residual correlation in the error terms. Specifically, 
\begin{align*}
M_1 &= \alpha_0 + \alpha_1A_1 + \boldsymbol{\alpha_2'C} + \epsilon_1 \\
M_2 &= \gamma_0 + \gamma_1A_2 + \boldsymbol{\gamma_2'C} + \epsilon_2 
\end{align*}
where $\big(
\begin{smallmatrix}
\epsilon_1 \\ \epsilon_2
\end{smallmatrix}
\big) \sim MVN\left( \boldsymbol{\mu} = 
\big(\begin{smallmatrix}
0 \\ 0
\end{smallmatrix}\big), \boldsymbol{\Sigma} = \big(\begin{smallmatrix}
\sigma_1^2 &\rho\sigma_1\sigma_2 \\
\rho\sigma_1\sigma_2 & \sigma_2^2
\end{smallmatrix}\big)
\right).$ Here, $\rho$ represents the Pearson correlation coefficient between $\epsilon_1$ and $\epsilon_2$. In the model fit to the observed data $A_1 = A_2 = A$.

If both $M_1$ and $M_2$ represent binary mediator variables, we utilize a bivariate probit model to jointly model the correlated mediators. Under this model, we assume that each mediator variable arises from an underlying continuous random variable. Furthermore, the correlation is in the error terms of the underlying continuous random variables. Let $M_1^*$ and $M_2^*$ denote the underlying continuous random variables corresponding to $M_1$ and $M_2$, respectively. To connect these underlying continuous random variables to the observed binary variables, we let $M_j = 1$ if $M_j^* > 0$ and $M_j = 0$ otherwise for $j = 1,2$. The bivariate probit mediator model is specified as:
\begin{align*}
M_1^* &= \alpha_0 + \alpha_1A_1 + \boldsymbol{\alpha_2'C} + \epsilon_1, \hspace{0.3cm} M_1 = I(M_1^* > 0) \\
M_2^* &= \gamma_0 + \gamma_1A_2 + \boldsymbol{\gamma_2'C} + \epsilon_2, \hspace{0.3cm} M_2 = I(M_2^* > 0) 
\end{align*}
where $\big(
\begin{smallmatrix}
\epsilon_1 \\ \epsilon_2
\end{smallmatrix}
\big) \sim MVN\left( \boldsymbol{\mu} = 
\big(\begin{smallmatrix}
0 \\ 0
\end{smallmatrix}\big), \boldsymbol{\Sigma} = \big(\begin{smallmatrix}
1 &\rho \\
\rho &1
\end{smallmatrix}\big)
\right).$
Any combination of binary and continuous mediator variables may also be expressed in terms of a joint multivariate linear regression/ probit model. If $M_1$ were binary and $M_2$ were continuous, the following model formulation would capture correlation in $M_1$ and $M_2$ beyond the correlation that is accounted for by including $A$ and $\boldsymbol{C}$ in the model:
\begin{align*}
M_1^* &= \alpha_0 + \alpha_1A_1 + \boldsymbol{\alpha_2'C} + \epsilon_1, \hspace{0.3cm} M_1 = I(M_1^* > 0) \\
M_2 &= \gamma_0 + \gamma_1A_2 + \boldsymbol{\gamma_2'C} + \epsilon_2
\end{align*}
where $\big(
\begin{smallmatrix}
\epsilon_1 \\ \epsilon_2
\end{smallmatrix}
\big) \sim MVN\left( \boldsymbol{\mu} = 
\big(\begin{smallmatrix}
0 \\ 0
\end{smallmatrix}\big), \boldsymbol{\Sigma} = \big(\begin{smallmatrix}
1 &\rho\sigma_2 \\
\rho\sigma_2 & \sigma_2^2
\end{smallmatrix}\big)
\right).$
While we focus on the setting of two correlated mediators, note that this approach can be used for more than two mediator variables by expanding the error covariance matrix to include all pairwise correlations between the jointly modelled mediators. If three correlated mediator variables were of interest in a causal decomposition analysis, $\boldsymbol{\Sigma}$ would be parametrized by $\sigma_1^2$, $\sigma_2^2$, $\sigma_3^2$, $\rho_{12}$, $\rho_{13}$, and $\rho_{23}$, where $\rho_{jk}$ represents the Pearson correlation between the error terms corresponding to mediators $j$ and $k$.

\subsection{Causal assumptions} \label{sec:assumptions}
Several assumptions must hold to draw causal conclusions from this decomposition analysis. First, by the consistency assumption, the counterfactual outcomes are connected to the observed data values and models. $Y(A, \boldsymbol{m}) = Y$ when $\boldsymbol{M}$ is set to $\boldsymbol{m}$ and the group is held at its observed value. Under the conditional exchangeability assumption, $Y(A, m_1, m_2) \perp M_1, M_2 \mid \boldsymbol{C}$ for all $m_1, m_2$. Broadly, this assumption suggests that there is no unmeasured confounding in the mediator-outcome relationship. In this assumption, we are implicitly conditioning on $A$, and as discussed earlier, $A$ must not be a collider on the directed acyclic graph (DAG). We also assume that the parametric outcome and mediator models are correctly specified. The proposed joint mediator model assumes that error terms for the mediator variables or latent mediator variables follow a multivariate normal distribution, and that this distribution adequately corrects for the residual correlation. Finally, and perhaps the assumption that must be made most carefully, is the assumption that the covariance matrix dictating the residual correlation between the mediators is constant, regardless of the value of $A$. A similar assumption is made in the mediation analysis paper by \cite{wang2013estimation}. This assumption is necessary for identifying the path-specific decomposition effects.

\subsection{Causal decomposition effects}
Like the method proposed by \cite{sudharsanan2021educational}, our extension is highly flexible. Virtually any summary measure of the health disparity between groups may be employed. With a binary outcome, examples include a relative risk or risk difference between groups $A=1$ and $A=0$ in each pseudopopulation. In addition, any metric to compare these summary measures between the natural-course and counterfactual pseudopopulations may be selected. In this paper, we use the set of effects detailed in Table \ref{tab:effect_definitions} to summarize the results of our decomposition analysis.
\begin{table}[ht]
\caption{Effects that summarize the difference in health outcome between groups under the natural and counterfactual scenarios. The effects of interventions on specific mediators, or path-specific effects, are included in addition to the joint effect on $M_1$ and $M_2$ together.}
\label{tab:effect_definitions}
\resizebox{\textwidth}{!}{%
\begin{tabular}{|l|l|l|l|}
\hline
\textbf{Effect}                                                                      & \textbf{Notation} & \textbf{Expression}                                                                                                                                         & \textbf{Description}                                                                                                                                                                                                                                                \\ \hline
Natural-course RR                                                                    & $RR_{natural}$                                                         & $\frac{E(Y(1, \boldsymbol{M} \sim f_{\boldsymbol{M}\mid 1, 1, \boldsymbol{C}}))}{E(Y(0, \boldsymbol{M} \sim f_{\boldsymbol{M}\mid 0, 0, \boldsymbol{C}}))}$ & \begin{tabular}[c]{@{}l@{}}Relative risk of $Y=1$ ($A=1$ vs. $A = 0$)\\ in the natural-course pseudopopulation\end{tabular}                                                                                                                                         \\ \hline
\begin{tabular}[c]{@{}l@{}}Counterfactual RR - \\ joint intervention\end{tabular}    & $RR_{count,0,0}$                                                       & $\frac{E(Y(1, \boldsymbol{M} \sim f_{\boldsymbol{M}\mid 0, 0, \boldsymbol{C}}))}{E(Y(0, \boldsymbol{M} \sim f_{\boldsymbol{M}\mid 0, 0, \boldsymbol{C}}))}$ & \begin{tabular}[c]{@{}l@{}}Relative risk of $Y=1$ ($A=1$ vs. $A=0$)\\ in the counterfactual pseudopopulation \\ where a joint intervention is employed to \\ equalize the distribution of $M_1$ and $M_2$ \\ in group $A = 1$ to that of group $A = 0$\end{tabular} \\ \hline
\begin{tabular}[c]{@{}l@{}}Counterfactual RR - \\ intervention on $M_1$\end{tabular} & $RR_{count,0,1}$                                                       & $\frac{E(Y(1, \boldsymbol{M} \sim f_{\boldsymbol{M}\mid 0, 1, \boldsymbol{C}}))}{E(Y(0, \boldsymbol{M} \sim f_{\boldsymbol{M}\mid 0, 0, \boldsymbol{C}}))}$ & \begin{tabular}[c]{@{}l@{}}Relative risk of $Y=1$ ($A=1$ vs. $A=0$)\\ in the counterfactual pseudopopulation \\ where an intervention is employed to \\ equalize the distribution of $M_1$ \\ in group $A = 1$ to that of group $A = 0$\end{tabular}                \\ \hline
\begin{tabular}[c]{@{}l@{}}Counterfactual RR - \\ intervention on $M_2$\end{tabular} & $RR_{count,1,0}$                                                       & $\frac{E(Y(1, \boldsymbol{M} \sim f_{\boldsymbol{M}\mid 1, 0, \boldsymbol{C}}))}{E(Y(0, \boldsymbol{M} \sim f_{\boldsymbol{M}\mid 0, 0, \boldsymbol{C}}))}$ & \begin{tabular}[c]{@{}l@{}}Relative risk of $Y=1$ ($A=1$ vs. $A=0$)\\ in the counterfactual pseudopopulation \\ where an intervention is employed to \\ equalize the distribution of $M_2$ \\ in group $A = 1$ to that of group $A = 0$\end{tabular}                \\ \hline
\begin{tabular}[c]{@{}l@{}}Reduction in RR - \\ joint intervention\end{tabular}      & $RR_{red,0,0}$                                                         & $RR_{natural} - RR_{count,0,0}$                                                                                                                             & Joint intervention effect                                                                                                                                                                                                                                           \\ \hline
\begin{tabular}[c]{@{}l@{}}Reduction in RR - \\ intervention on $M_1$\end{tabular}   & $RR_{red,0,1}$                                                         & $RR_{natural} - RR_{count,0,1}$                                                                                                                             & \begin{tabular}[c]{@{}l@{}}Path-specific intervention effect\\ through $M_1$\end{tabular}                                                                                                                                                                           \\ \hline
\begin{tabular}[c]{@{}l@{}}Reduction in RR - \\ intervention on $M_2$\end{tabular}   & $RR_{red,1,0}$                                                         & $RR_{natural} - RR_{count,1,0}$                                                                                                                             & \begin{tabular}[c]{@{}l@{}}Path-specific intervention effect \\ through $M_2$\end{tabular}                                                                                                                                                                          \\ \hline
\end{tabular}%
}
\end{table}

\subsection{Algorithm and implementation} \label{sec:algorithm}
To perform our decomposition analysis, we adapt the Monte Carlo integration-based decomposition algorithm proposed by \cite{sudharsanan2021educational}. Importantly, we describe how to implement this algorithm in the setting of multiple, correlated mediators using the joint mediator model and causal assumptions specified in Sections \ref{sec:joint} and \ref{sec:assumptions}, respectively. We highlight the key function needed to implement this method in R for two mediators that may be any combination of binary and continuous variables. The full code is available in the appendix. Under the intervention which equalizes the mediator distributions in group $A = 1$ to the mediator distributions in group $A = 0$, both individually and jointly, the steps are provided below:

\begin{enumerate}
\item Fit the outcome model to the observed data.
\item Fit the joint mediator model to the observed data. To fit the proposed joint mediator model in R, use the \texttt{gjrm} function from the \texttt{GJRM} package \citep{marra2017joint} as follows:
\begin{verbatim}
  med_model <- gjrm(list(m1 ~ a1 + c, m2 ~ a2 + c),
                    data = data, 
                    margins = c("probit", "N"), 
                    Model = "B")
\end{verbatim}
In the code above, the \texttt{margins} argument indicates that the first mediator is binary (denoted as \texttt{"probit"}), and the second mediator is normally distributed (denoted as \texttt{"N"}). If the mediators are both continuous, \texttt{margins} should be set to \texttt{c("N", "N")} and if the mediators are both binary, \texttt{margins} should be set to \texttt{c("probit", "probit")}.
\item Extract the estimated variance parameters and correlation coefficients of the error terms from the joint mediator model fit. Then, construct an estimated covariance matrix for the error terms using these components. 
\item Create a natural-course pseudopopulation. To do this, plug the observed value of $A$ and $\boldsymbol{C}$ into the fitted joint mediator model and simulate a large number ($K$) of ``natural'' mediator values (or latent mediator values, for binary mediators) for each individual using the estimated covariance matrix to simulate the correlated random error terms. If a mediator is binary, compute the simulated mediator value from the latent mediator value using the indicator function. Then, plug these simulated natural mediator variables along with the observed value of $A$ and $\boldsymbol{C}$ into the fitted outcome model to simulate $K$ natural outcome values for each individual.
\item From each of the $K$ sets of outcomes, compute the contrast of interest in the decomposition analysis between groups $A = 1$ and $A = 0$. If the contrast of interest is the relative risk, this would be computed as follows for the $k$th simulated values:
$$RR_{natural}^{(k)} =\frac{\frac{1}{N_1}\sum_{i = 1}^N Y_i^{(k)}(1, \boldsymbol{M}^{(k)}_i \sim f_{\boldsymbol{M}\mid 1, 1, \boldsymbol{C}_i})*I(A_i = 1)}{\frac{1}{N_0}\sum_{i = 1}^N Y_i^{(k)}(0, \boldsymbol{M}^{(k)}_i \sim f_{\boldsymbol{M}\mid 0, 0, \boldsymbol{C}_i})*I(A_i = 0)}$$ 
where $I$ denotes the indicator function, $N_1$ is the number of individuals in group $A = 1$, and $N_0$ is the number of individuals in group $A = 0$. Average all $K$ values of this contrast to obtain an estimate with minimal Monte Carlo error.
\item Compute the contrast under each counterfactual scenario to create counterfactual pseudopopulations corresponding to each intervention. 

For the intervention that equalizes the distribution of all mediators in group $A = 1$ to that of group $A = 0$, simulate values from the joint distribution of $\boldsymbol{M}$, where $A_1$ and $A_2$ are both set to $0$ for all individuals. Then, plug in the sampled mediator values to obtain counterfactual outcome values for each individual. The counterfactual relative risk in the $k$th simulated dataset can be computed as:
$$RR_{count,0,0}^{(k)} =\frac{\frac{1}{N_1}\sum_{i = 1}^N Y_i^{(k)}(1, \boldsymbol{M}^{(k)}_i \sim f_{\boldsymbol{M}\mid 0, 0, \boldsymbol{C}_i})*I(A_i = 1)}{\frac{1}{N_0}\sum_{i = 1}^N Y_i^{(k)}(0, \boldsymbol{M}^{(k)}_i \sim f_{\boldsymbol{M}\mid 0, 0, \boldsymbol{C}_i})*I(A_i = 0)}$$
Similarly, the counterfactual relative risks when only one mediator is equalized to group $A = 0$ are calculated from the $k$th samples as follows:
$$RR_{count,0,1}^{(k)} =\frac{\frac{1}{N_1}\sum_{i = 1}^N Y_i^{(k)}(1, \boldsymbol{M}^{(k)}_i \sim f_{\boldsymbol{M}\mid 0, 1, \boldsymbol{C}_i})*I(A_i = 1)}{\frac{1}{N_0}\sum_{i = 1}^N Y_i^{(k)}(0, \boldsymbol{M}^{(k)}_i \sim f_{\boldsymbol{M}\mid 0, 0, \boldsymbol{C}_i})*I(A_i = 0)}$$ 
$$RR_{count,1,0}^{(k)} =\frac{\frac{1}{N_1}\sum_{i = 1}^N Y_i^{(k)}(1, \boldsymbol{M}^{(k)}_i \sim f_{\boldsymbol{M}\mid 1, 0, \boldsymbol{C}_i})*I(A_i = 1)}{\frac{1}{N_0}\sum_{i = 1}^N Y_i^{(k)}(0, \boldsymbol{M}^{(k)}_i \sim f_{\boldsymbol{M}\mid 0, 0, \boldsymbol{C}_i})*I(A_i = 0)}$$ 
Average all $K$ values obtained for each of these effects.
\item Use each of these effects to compute a point estimate for the path-specific intervention effects: $RR_{red,0,0}$, $RR_{red,0,1}$, and $RR_{red,1,0}$.
\item Use a bootstrapping procedure to obtain confidence intervals for each decomposition effect.
\end{enumerate}

\section{Simulation study}
\label{section:simulation}
In this section, we compare the performance of our proposed method to the existing method proposed by \cite{sudharsanan2021educational} and implemented in \cite{sudharsanan2020rural} in the setting of multiple mediator variables. We compare the two methods under different levels of residual correlation, different types of mediators (binary, continuous, or mixed), and in the presence or absence of a mediator-mediator interaction effect on the health outcome.

\subsection{Simulation set-up}

To evaluate the proposed and existing methods, we simulated $1000$ datasets of size $n = 500$ under each scenario. In this simulation study, we focused on three metrics: percent bias, average 95\% confidence interval width, and 95\% confidence interval coverage for each decomposition effect (joint and path-specific) in a two-mediator setting. 95\% confidence intervals were obtained using $200$ bootstrap samples. Since both the proposed and existing methods utilize Monte Carlo methods, all estimates were based on $K = 500$ repetitions of the sampling steps.

\begin{figure}
\begin{center}
\includegraphics{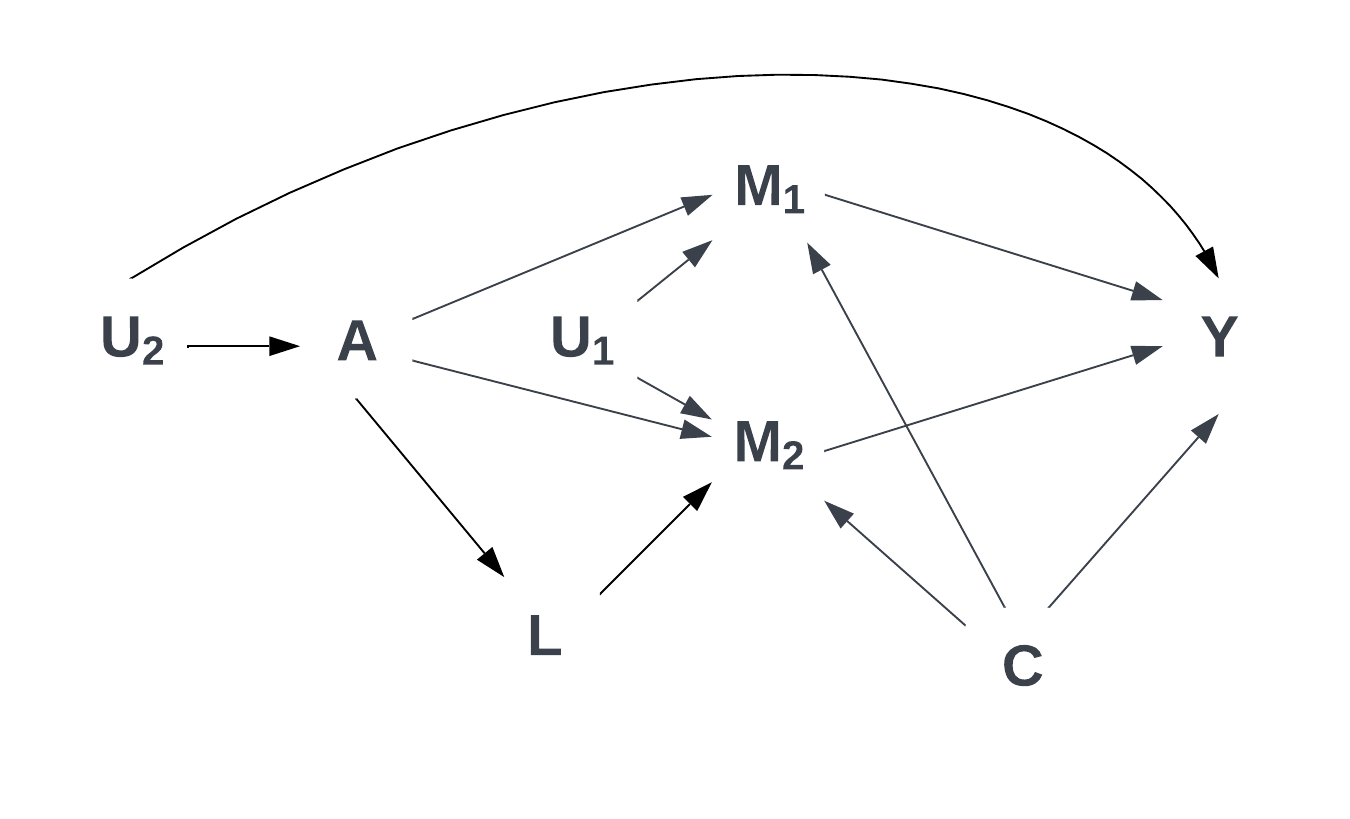}
\caption{DAG that depicts the assumed relationships in the simulation study. Variables $U_1$ and $U_2$ represent unobserved variables. $U_1$ induces the correlation between $M_1$ and $M_2$ while $U_2$ creates a non-causal pathway between $A$ and $Y$.}
\label{fig:simulation_dag}
\end{center}
\end{figure}

The data were generated according to the DAG in Figure \ref{fig:simulation_dag}. This DAG assumes that unmeasured variable $U_1$ induces a correlation between $M_1$ and $M_2$ and unmeasured variable $U_2$ creates a backdoor path between $A$ and $Y$, respectively. We also included an indirect pathway between $A$ and $M_2$ through the variable $L$ to demonstrate one of many underlying scenarios that a causal decomposition analysis can address.

To generate $C$, $U_1$, $U_2$, $A$, and $L$, we sampled from the following distributions:
\begin{itemize}
    \item $C \sim \text{Bernoulli}(\pi = 0.5)$
    \item $U_1 \sim \text{Normal}(\mu = 0, \sigma^2 = 1)$
    \item $U_2 \sim \text{Bernoulli}(\pi = 0.5)$
    \item $A \sim \text{Bernoulli}(\pi)$ where $\log\left(\frac{\pi}{1-\pi}\right) = 0 + 2U_2$
    \item $L \sim \text{Normal}(\mu, \sigma^2 = 1)$ where $\mu = 0 + 2A$
\end{itemize}
Using these samples, we then generated the mediator variables. To do this, we first generated the error terms for the continuous or underlying continuous random variables as $\epsilon_j \sim \text{Normal}(\mu = 0, \sigma^2 = 1)$ for $j = 1,2$. We computed the linear predictors ($LP_j$) for each mediator variable as:
\begin{align*}
LP_1 &= -1 + 0.5A + 0.5C + \theta_1U_1\\
LP_2 &= -1 + 0.5A + 0.5C + 0.5L + \theta_2U_1
\end{align*}

For the data-generating scenarios with two continuous mediators, $M_1$ and $M_2$ were then computed as $M_1 = LP_1 + \epsilon_1$ and $M_2 = LP_2 + \epsilon_2$. In the data-generating scenarios with one continuous and one binary mediator, $M_1$ was generated from a probit model as $M_1 = I(LP_1 + \epsilon_1 > 0)$ and $M_2$ was generated from a linear regression model as $M_2 = LP_2 + \epsilon_2$. Finally, with two binary mediators, $M_1 = I(LP_1 + \epsilon_1 > 0)$ and $M_2 = I(LP_2 + \epsilon_2 > 0)$.

After generating the mediator variables, which exibited residual correlation if $\theta_1$ and $\theta_2$ were non-zero, we generated the binary outcome variable using the model:
\begin{align*}
Y &\sim \text{Bernoulli}(\pi) \\
\log\left(\frac{\pi}{1-\pi}\right) &= -2 + 0.5U_2 + 0.5M_1 + 0.5M_2 + \theta_3M_1M_2 + 0.5C
\end{align*}

Across simulation scenarios, we varied the amount of residual correlation in the mediator values by changing $\theta_1$ and $\theta_2$, the coefficients corresponding to the unmeasured confounder. To evaluate the effect of a mediator-mediator interaction, we set $\theta_3 = 0.5$ in the scenarios with an interaction and $\theta_3 = 0$ in the scenarios without an interaction. Table \ref{tab:scenarios} provides a complete description of each simulation scenario including the values for $\theta_1$, $\theta_2$, and $\theta_3$.

The remaining true coefficient values were selected so that the true natural-course relative risk values were between 1.25 and 2.50 and the true counterfactual relative risks under the joint intervention fell between 1.10 and 1.25. Furthermore, in all scenarios, $M_2$ was simulated to have a stronger relationship with $A$ than $M_1$.

\begin{table}[ht]
\caption{Scenarios considered in the simulation study. Residual correlation indicates the approximate Pearson correlation between the error terms in the mediator models when excluding the unmeasured confounder, $U_1$.}
\label{tab:scenarios}
\resizebox{\textwidth}{!}{%
\begin{tabular}{|l|l|l|l|l|}
\hline
\textbf{Scenario} & \textbf{Mediator types}                                               & \textbf{Residual correlation} & \textbf{\begin{tabular}[c]{@{}l@{}}Mediator-mediator \\ interaction\end{tabular}} & \textbf{$\boldsymbol{(\theta_1, \theta_2, \theta_3)'}$} \\ \hline
1                 & Two continuous                                                        & 0                             & No                                                                                & $(0,0,0)'$                                              \\ \hline
2                 & Two continuous                                                        & 0.3                           & No                                                                                & $(0.7,0.7,0)'$                                          \\ \hline
3                 & Two continuous                                                        & 0.6                           & No                                                                                & $(1.25,1.25,0)'$                                        \\ \hline
4                 & Two continuous                                                        & 0                             & Yes                                                                               & $(0,0,0.5)'$                                            \\ \hline
5                 & Two continuous                                                        & 0.3                           & Yes                                                                               & $(0.7,0.7,0.5)'$                                        \\ \hline
6                 & Two continuous                                                        & 0.6                           & Yes                                                                               & $(1.25,1.25,0.5)'$                                      \\ \hline
7                 & \begin{tabular}[c]{@{}l@{}}One binary, \\ one continuous\end{tabular} & 0                             & No                                                                                & $(0,0,0)'$                                              \\ \hline
8                 & \begin{tabular}[c]{@{}l@{}}One binary, \\ one continuous\end{tabular} & 0.3                           & No                                                                                & $(0.7,0.7,0)'$                                          \\ \hline
9                 & \begin{tabular}[c]{@{}l@{}}One binary, \\ one continuous\end{tabular} & 0.6                           & No                                                                                & $(1.25,1.25,0)'$                                        \\ \hline
10                & \begin{tabular}[c]{@{}l@{}}One binary, \\ one continuous\end{tabular} & 0                             & Yes                                                                               & $(0,0,0.5)'$                                            \\ \hline
11                & \begin{tabular}[c]{@{}l@{}}One binary, \\ one continuous\end{tabular} & 0.3                           & Yes                                                                               & $(0.7,0.7,0.5)'$                                        \\ \hline
12                & \begin{tabular}[c]{@{}l@{}}One binary, \\ one continuous\end{tabular} & 0.6                           & Yes                                                                               & $(1.25,1.25,0.5)'$                                      \\ \hline
13                & Two binary                                                            & 0                             & No                                                                                & $(0,0,0)'$                                              \\ \hline
14                & Two binary                                                            & 0.3                           & No                                                                                & $(0.7,0.7,0)'$                                          \\ \hline
15                & Two binary                                                            & 0.6                           & No                                                                                & $(1.25,1.25,0)'$                                        \\ \hline
16                & Two binary                                                            & 0                             & Yes                                                                               & $(0,0,0.5)'$                                            \\ \hline
17                & Two binary                                                            & 0.3                           & Yes                                                                               & $(0.7,0.7,0.5)'$                                        \\ \hline
18                & Two binary                                                            & 0.6                           & Yes                                                                               & $(1.25,1.25,0.5)'$                                      \\ \hline
\end{tabular}%
}
\end{table}

Finally, in order to obtain estimates of the percent bias and confidence interval coverage, we computed the true values for each effect in the decomposition analysis. Because there was no closed-form solution for these effects, we obtained the true effects using Monte Carlo integration based on 100,000 samples. We further reduced error in the estimates of the true values by averaging the estimates obtained when performing this integration 100 times. Monte Carlo integration was used to approximate the following integral:
\begin{align*}
&E(Y(A, \boldsymbol{M} \sim f_{\boldsymbol{M}\mid a_1, a_2, c})) \\= &\int \dots \int E(y \mid m_1, m_2, c, u_2)dF(m_1\mid a_1, c, u_1)dF(m_2 \mid a_2, c, l, u_1)dF(l \mid a_2)dF(u_2 \mid a)dF(u_1)dF(c)
\end{align*}

By integrating this quantity under different values of $A = a$, $a_1$, and $a_2$, we were able to obtain precise approximations of the true decomposition effect. Note that this integral may change depending on the true relationships in the DAG and reflects the DAG in Figure \ref{fig:simulation_dag}. See the appendix for the full derivation of this integral.
\subsection{Simulation results}
\begin{figure}[ht]
    \centering
    \includegraphics[scale = 0.1]{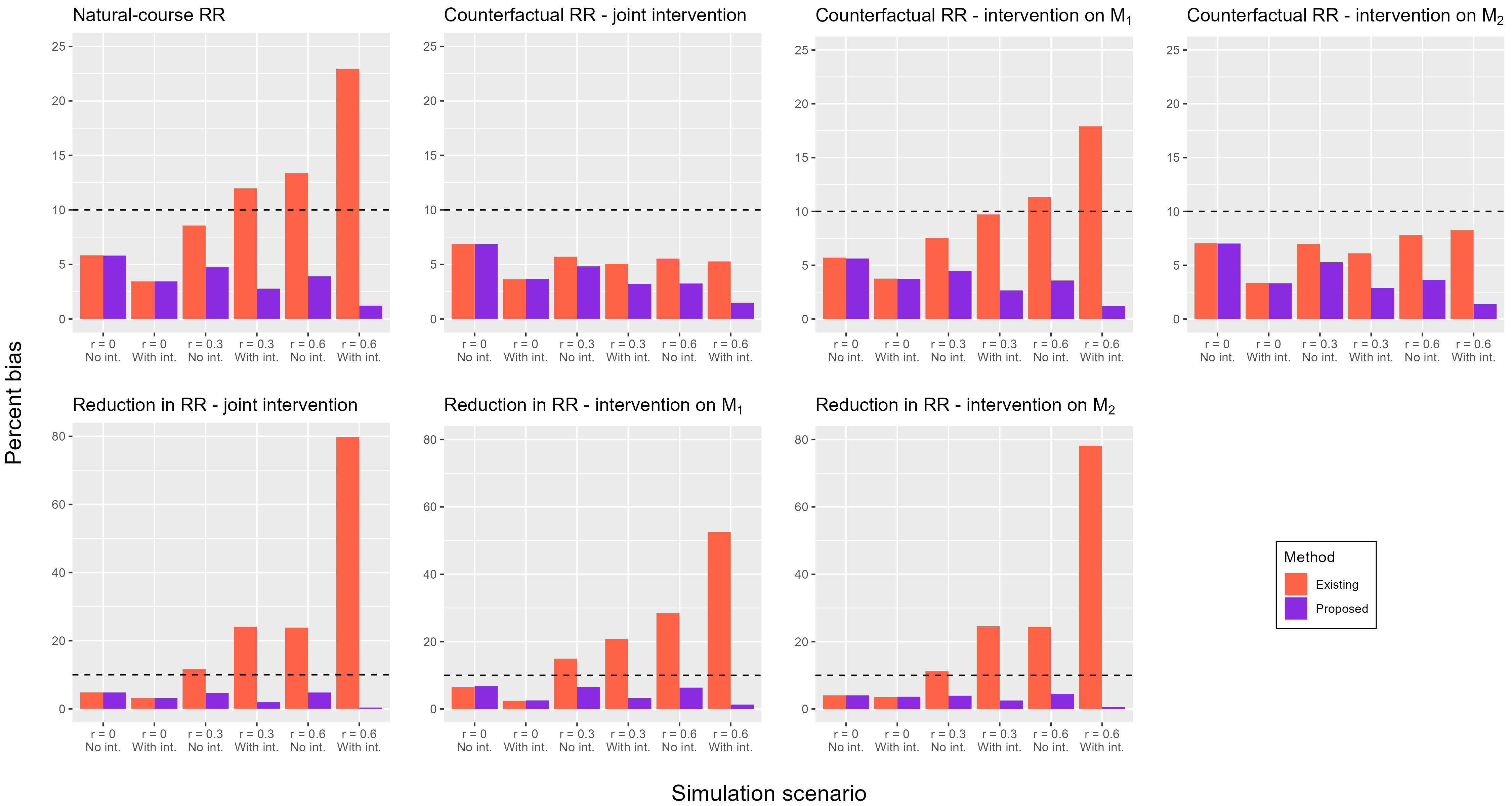}
    \caption{Percent bias for each decomposition effect when both mediators were continuous and the relative risk was used to summarize the health differences between groups. $r$ is the approximate residual correlation and `int' indicates a mediator-mediator interaction effect.}
    \label{fig:two_continuous_rr_bias}
\end{figure}

\begin{figure}[ht]
    \centering
    \includegraphics[scale = 0.1]{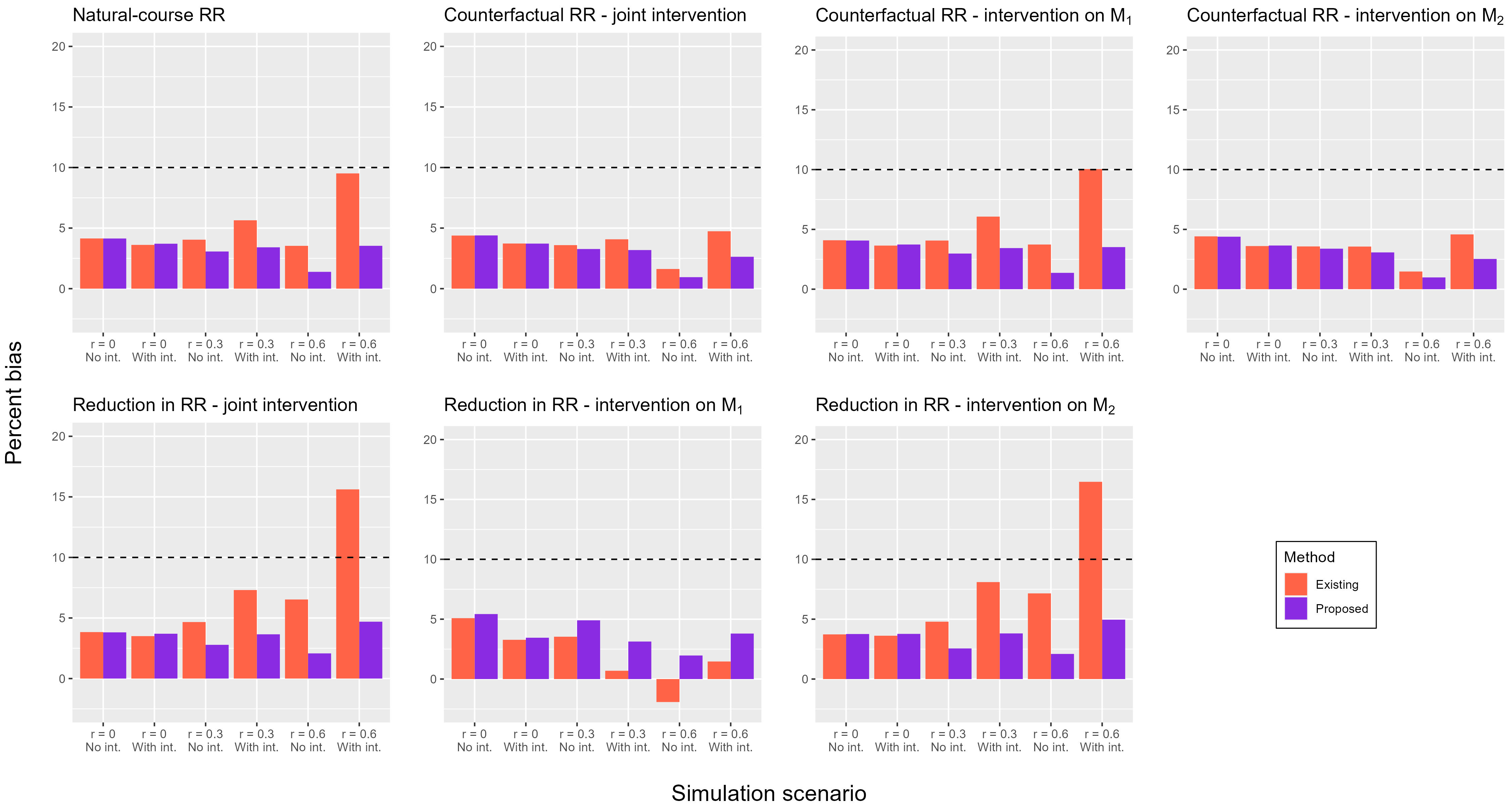}
    \caption{Percent bias for each decomposition effect when $M_1$ was binary and $M_2$ was continuous. Relative risk was used to summarize the health differences between groups. $r$ is the approximate residual correlation and `int' indicates a mediator-mediator interaction effect.}
\label{fig:mixed_rr_bias}
\end{figure}

\begin{figure}[ht]
    \centering
    \includegraphics[scale = 0.1]{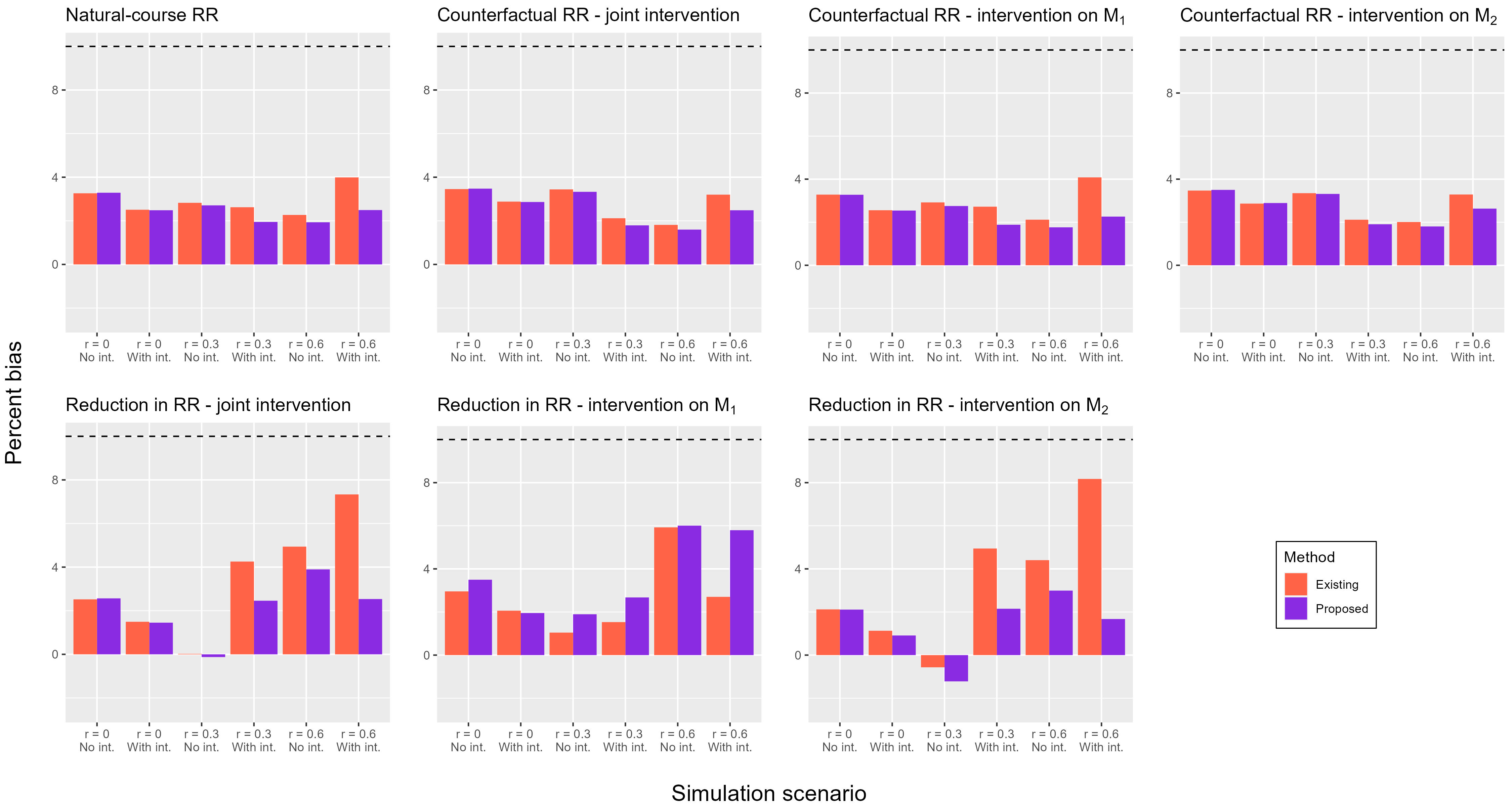}
    \caption{Percent bias for each decomposition effect when both mediators were binary and the relative risk was used to summarize the health differences between groups. $r$ is the approximate residual correlation and `int' indicates a mediator-mediator interaction effect.}
\label{fig:two_binary_rr_bias}\end{figure}

Figures \ref{fig:two_continuous_rr_bias}, \ref{fig:mixed_rr_bias}, and \ref{fig:two_binary_rr_bias} display the percent bias in each of the decomposition effects computed for two continuous mediators, mixed mediators, and two binary mediators, respectively. In each figure of the plot grids, a dotted line is depicted at 10\% and also at -10\% if bias occurs in the negative direction. The bottom rows of the plot grids display the joint intervention effect followed by the two path-specific effects.
Figure \ref{fig:two_continuous_rr_bias} demonstrates a substantial improvement in bias using the proposed method compared to the existing method when two continuous mediators were included, particularly as the residual correlation increased. When the residual correlation was approximately 0, there was no difference in bias when applying either method. However, when the residual bias increased to 0.3, bias in the joint intervention effect and path-specific intervention effects exceeded 10\% under the existing method yet remained close to 5\% or lower using the proposed method. When the residual correlation was approximately 0.6, the joint and path-specific effects obtained using the existing approach demonstrated substantial bias, and the presence of the mediator-mediator interaction further exacerbated this bias. The percent bias in the scenario with a residual correlation of approximately 0.6 and a mediator-mediator interaction effect was 80\%, 53\%, and 78\% for the joint effect, path-specific effect through $M_1$, and path-specific effect through $M_2$ under the existing method, whereas the proposed method resulted in percent biases of under 1.5\% for each of these quantities. These results suggest that in terms of bias, use of the proposed method is highly advantageous over use of the existing method when two moderately-correlated continuous mediators are considered.

The percent biases for the scenarios involving one continuous and one binary mediator are displayed in Figure \ref{fig:mixed_rr_bias}. In each of these scenarios, the bias in decomposition effects was less pronounced than in the scenarios involving two continuous mediators. Once again, the existing method resulted in more biased estimates than the proposed method for most of the decomposition effects at residual correlation levels of 0.3 and 0.6. Interestingly, though, the existing method exhibited improved bias for the path-specific effect for $M_1$, while the proposed method exhibited improved bias for the path-specific effect for $M_2$. This is likely due to the small size of the path-specific decomposition effect through $M_1$ as compared to the size of the effect through $M_2$. Given that the percent bias always hovered around 5\% or lower under the proposed method, the higher bias for the path-specific effect through $M_1$ is not particularly concerning. Overall, the proposed method resulted in an improved performance over the existing method in the case of mixed mediator types, where one was binary and the other was continuous.

Figure \ref{fig:two_binary_rr_bias} displays the biases for each decomposition effect in the scenarios involving two binary mediator variables. Here, we can see that each method always produced effects with percent biases less than 10\%. Typically, the proposed method slightly improved on percent bias over the existing method, with the most noticeable differences in the joint effect and path-specific effects for $M_2$ with residual correlation levels of 0.6.
\begin{figure}[ht]
    \centering
    \includegraphics[scale = 0.1]{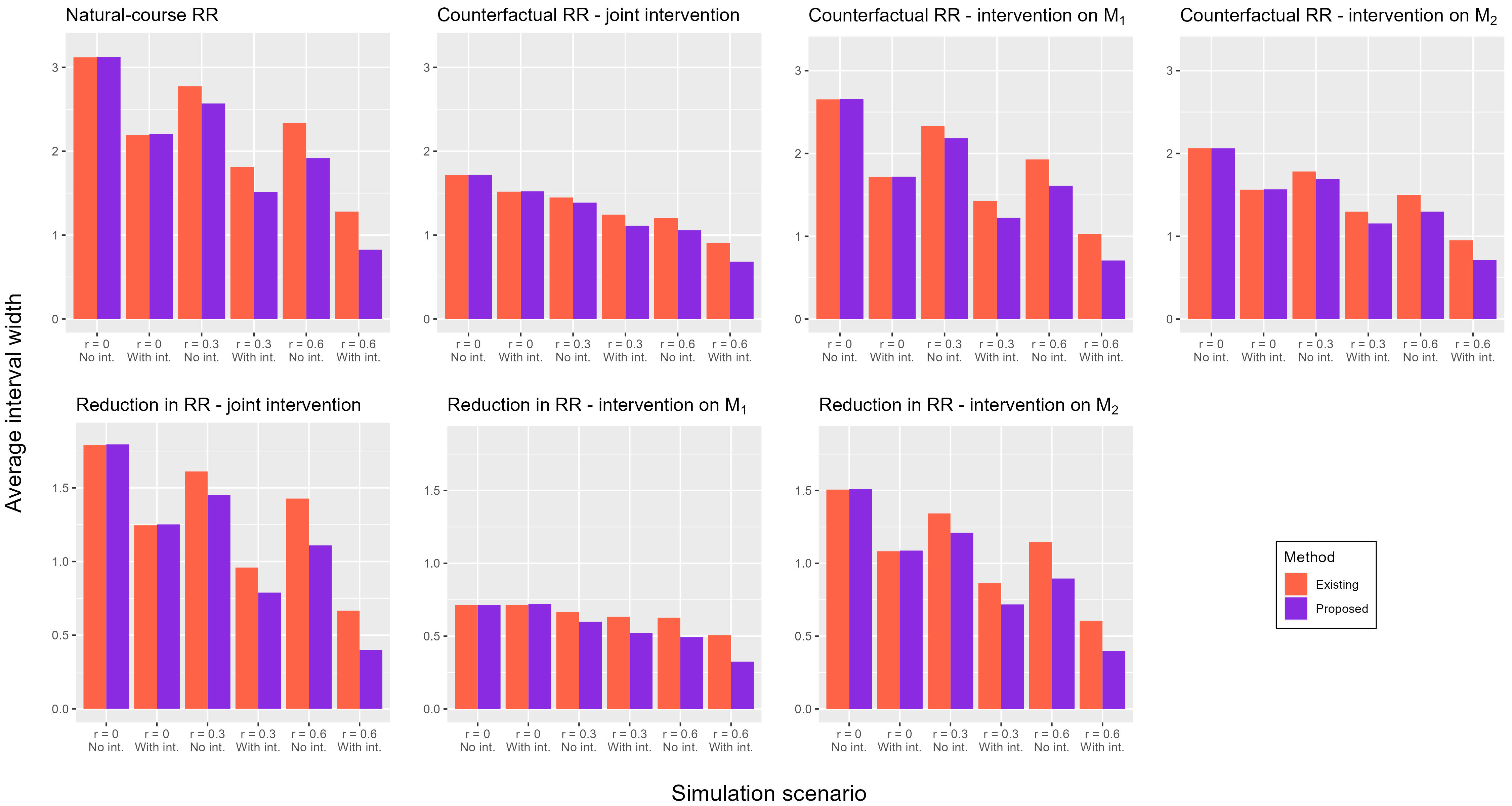}
    \caption{Average 95\% confidence interval width for each decomposition effect when both mediators were continuous and the relative risk was used to summarize the health differences between groups. $r$ is the approximate residual correlation and `int' indicates a mediator-mediator interaction effect.}
\label{fig:two_continuous_rr_interval_width}
\end{figure}

Next, we compared average 95\% confidence interval widths and coverage levels between the proposed and existing methods. The average confidence interval widths when both mediators are continuous are shown in Figure \ref{fig:two_continuous_rr_interval_width}. In each of the scenarios with two continuous mediators, the average confidence interval width was lower under the proposed method compared to the existing method in the presence of residual correlation. In the absence of residual correlation, each decomposition effect had similar average confidence interval widths under both methods, with the proposed method producing slightly wider intervals in some cases. There were very minimal differences in average confidence interval widths produced by each method in all of the simulation scenarios involving two binary mediators (shown in the appendix). The differences in average confidence interval widths under each method in the mixed mediators setting were less pronounced than in Figure \ref{fig:two_continuous_rr_interval_width} but still generally illustrated some benefit in using the proposed method compared to the existing method. The difference in the two methods' performance on average interval width did not vary much by the presence or absence of a mediator-mediator interaction. The figures corresponding to the average interval width in the binary and mixed mediator settings are available in the appendix.
\begin{figure}[ht]
    \centering
    \includegraphics[scale = 0.1]{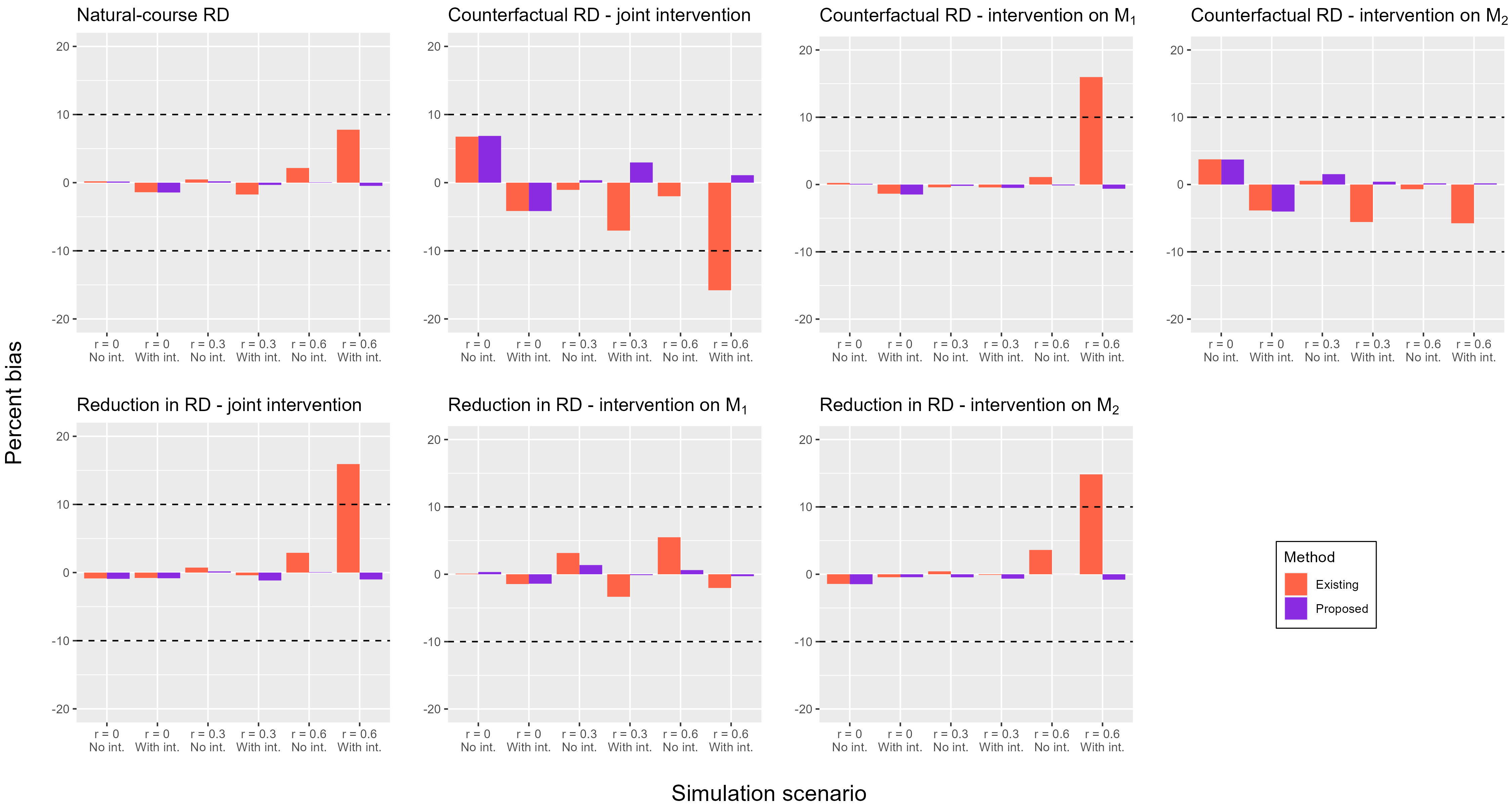}
    \caption{Percent bias for each decomposition effect when both mediators were continuous and the risk difference was used to summarize the health differences between groups. $r$ is the approximate residual correlation and `int' indicates a mediator-mediator interaction effect.}
\label{fig:two_continuous_rd_bias}
\end{figure}
Across all scenarios involving mixed or two binary mediators, coverage levels were similar between the proposed and existing methods, falling between 92\% and 96\%. Consistent with the earlier results related to bias and average interval width, the most notable differences in coverage occurred for the joint and path-specific effects where two continuous mediators with a residual correlated of 0.6 and an interaction effect were considered. In this scenario, the coverage for the joint intervention effect was 51.8\% under the existing method whereas it was 93.7\% under the proposed method. The path-specific effects computed from the existing method similarly had poor coverage at 85\% for the effect through $M_1$ and 54.7\% for the effect through $M_2$. The greatest improvements in terms of coverage between the proposed and existing methods were seen for two moderately correlated continuous mediators that exhibited an interaction effect on the outcome.

While not the focus of this simulation study, we also aimed to understand how much the bias results changed with a different choice of summary measure in health outcome between groups. As highlighted previously, both the proposed and existing methods are highly flexible and can be used with virtually any summary measure and contrast. Rather than computing the decomposition effects on the relative risk scale, we ran simulation scenarios that focused on the risk difference. Figure \ref{fig:two_continuous_rd_bias} displays the percent bias when two continuous mediators were of interest, but the risk difference was used to compare the health outcome in groups $A = 1$ vs. $A = 0$. Interestingly, the proposed method almost always performed better than the existing method, but the percent biases were much lower than on the relative risk scale. Nearly all of the percent biases under both methods fell well below 10\%. Recall that on the relative risk scale, the existing method produced joint and path-specific effects that were 50-80\% biased under the highest correlation scenarios. Thus, the choice of summary measure will have an effect on the bias in decomposition estimates. Our conclusions focus on the relative risk, a commonly-used summary measure in the literature.
\section{Application to racial differences in diabetes incidence in the REGARDS Study}
\label{section:application}
We applied our new methodology to understand whether differences in incident diabetes between Black and White adults in the United States could be reduced by intervening on two potential mediator variables, either individually or jointly: current smoking status (binary) and dietary inflammation score (DIS) (continuous).  DIS is a composite score made up of each individual's frequency of consumption of foods in 19 different food groups. DIS is associated with systemic inflammation, with higher scores indicating more exposure to proinflammatory diets \citep{byrd2019development}. We also applied the existing causal decomposition analysis method for comparison. This analysis used data from the Reasons for Geographic and Racial Differences in Stroke (REGARDS) Study \citep{howard2005reasons}.

REGARDS is a national prospective cohort study involving more than 30,000 White and Black adults \citep{howard2005reasons}. The REGARDS Study has made strides toward understanding factors that explain why Black adults are at a higher risk of stroke mortality than White adults and why there is a particularly high stroke mortality rate in the southern region of the United States \citep{howard2006racial, howard2011disparities, howard2017contributors}. In this study, demographic factors, information on stroke risk factors such as diet information, and physiologic measurements were collected at baseline. Participants were followed up by phone in six-month intervals to gather stroke outcome data, and a second in-home visit was completed approximately 10 years later. New diabetes diagnoses were recorded at the second in-home visit. For additional details on the REGARDS study design, see \cite{howard2005reasons}.

In a recent study by \cite{carson2021sex}, the authors performed a mediation analysis using the REGARDS dataset to assess which demographic, lifestyle, and clinical factors mediate the relationship between race and incident diabetes after stratifying by sex. Both smoking status and dietary variables were included in this mediation analysis. Because we expected there to be residual correlation among certain lifestyle variables and race is an intrinsic, non-manipulable characteristic, we illustrate the use of our proposed decomposition analysis method as an alternative approach to answer this research question. 500 Monte Carlo samples were drawn to compute each point estimate and 200 bootstrap intervals were used to construct confidence intervals.

As in the study by \cite{carson2021sex}, our outcome of interest was incident diabetes developed between the first and second in-home visits. All individuals with diabetes at the first in-home visit were excluded. Diabetes was defined as fasting glucose $\geq$ 126 mg/dL, random glucose $\geq$ 200 mg/dL, or use of oral or injectable hypoglycemic medications. The two groups of interest were Black adults and White adults, where the group of Black adults were coded as $A = 1$ and experienced a higher diabetes burden. The mediator-outcome confounders that were included in this analysis were prediabetes at baseline, education level, and median household income. The decomposition analyses were stratified by sex, since \cite{carson2021sex} illustrated differences in indirect effects by sex in their study. 

In total, 7902 individuals were included in this analysis. All of these individuals did not have diabetes at baseline and had complete data on the variables included in the decomposition analysis. 27.6\% of these individuals were Black, and 55.8\% were female. Diabetes incidence differed by race, with 9.5\% of the White adults developing diabetes and 17.5\% of the Black adults developing diabetes. The Black adults in the study population were more likely to report being a current smoker at baseline (14.5\%) compared to White adults (8.9\%). Furthermore, average DIS was higher in Black adults compared to White adults, with a mean DIS score of 0.72 and -0.71 for the two groups, respectively.

We fit the joint mediator model described in Section \ref{sec:joint} for one binary and one continuous mediator to these data. The estimated Pearson correlation coefficient for the error terms in the males' mediator model was $0.249$ with a 95\% confidence interval of (0.194, 0.308). The estimated Pearson correlation coefficient was quite similar in the model for females, at $0.231$ with a 95\% confidence interval of (0.192, 0.272). To verify the assumption required to identify path-specific effects, we examined the similarity between the estimated error covariance matrices between Black and White males and between Black and White females. Among males, the estimated covariances matrices for Black and White individuals were $\hat{\Sigma} = \big(\begin{smallmatrix}
1.00 & 0.64 \\
0.64 & 4.97
\end{smallmatrix}\big)$ and $\hat{\Sigma} = \big(\begin{smallmatrix}
1.00 & 0.54 \\
0.54 & 5.19
\end{smallmatrix}\big)$, which seemed fairly similar. Among females, these covariance matrices were $\hat{\Sigma} = \big(\begin{smallmatrix}
1.00 & 0.50 \\
0.50 & 5.48
\end{smallmatrix}\big)$ and $\hat{\Sigma} = \big(\begin{smallmatrix}
1.00 & 0.54 \\
0.54 & 4.96
\end{smallmatrix}\big)$, which also had minimal differences.
\begin{figure}[ht]
    \includegraphics[scale = 0.6]{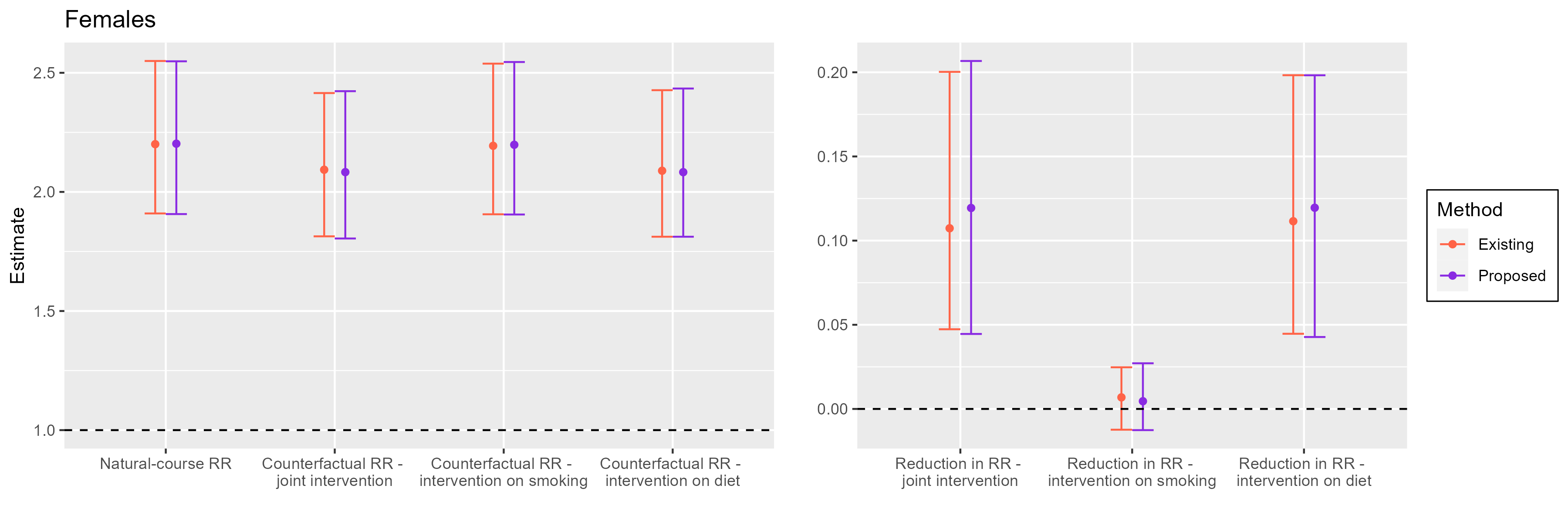}
\caption{Decomposition effect estimates and 95\% confidence intervals comparing diabetes incidence between Black and White females in the natural course and under joint and path-specific interventions.}
\label{fig:decomp_females}
\end{figure}

Figure \ref{fig:decomp_females} displays the decomposition effect estimates and 95\% confidence intervals comparing diabetes incidence between Black and White females in the REGARDS dataset. Under the proposed method, the natural-course relative risk was 2.20 (95\% CI: 1.91, 2.55) with a very similar estimate and 95\% confidence interval using the existing method. There was a statistically significant effect of the joint intervention and an intervention on just diet under both methods; intervening on smoking did not appear to narrow the gap in diabetes incidence between Black and White females under either method. Under the proposed method, the intervention on diet resulted in a reduction in RR of 0.12 (95\% CI: 0.04, 0.20), indicating that equalizing DIS in Black females to that of White females may modestly reduce the diabetes relative risk, but much of this difference would still be unexplained. The existing method produced similar conclusions, with small differences in the joint intervention effect and path-specific effect on the diabetes relative risk. The path-specific effect of diet on diabetes relative risk between Black and White females was estimated to be 7\% higher under the proposed method compared to the existing method; however, the absolute difference in the methods' estimates were quite small.
\begin{figure}[ht]
    \includegraphics[scale = 0.6]{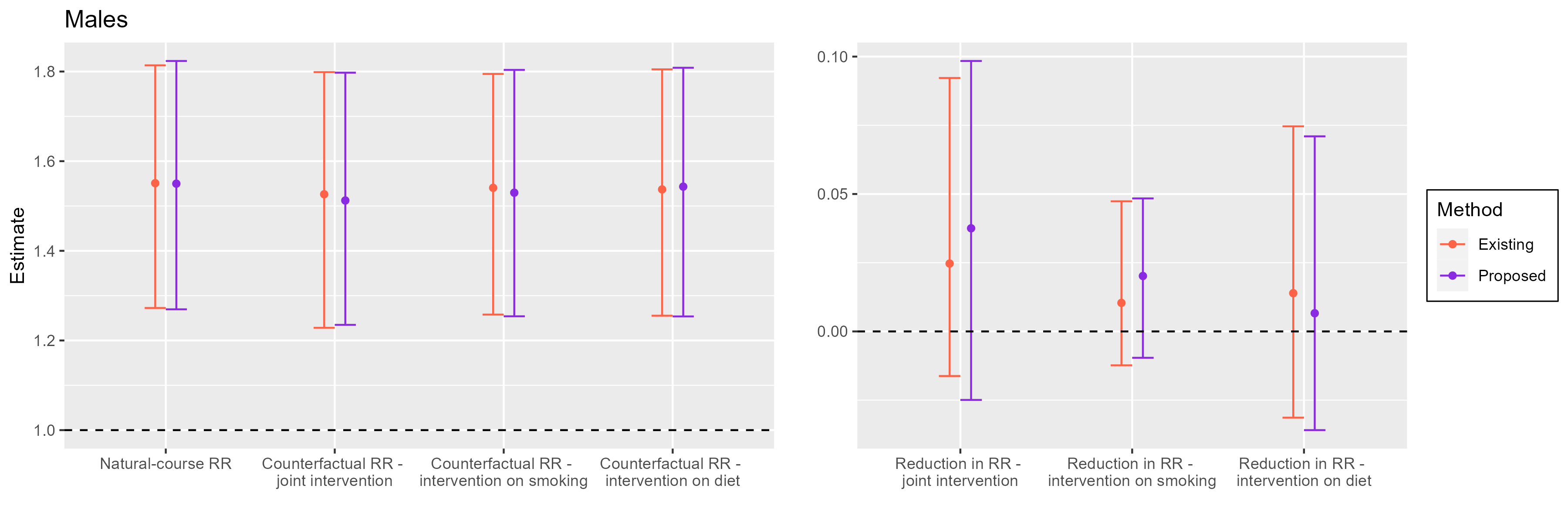}
\caption{Decomposition effect estimates and 95\% confidence intervals comparing diabetes incidence between Black and White males in the natural course and under joint and path-specific interventions.}
\label{fig:decomp_males}
\end{figure}

We similarly summarized the decomposition effects among males in Figure \ref{fig:decomp_males}. The natural-course relative risk estimated under the proposed method was lower than that of females at 1.55 (95\% CI: 1.27, 1.82). This effect was similarly summarized using the existing method. None of the interventions (on one or both mediators) resulted in statistically significant reductions in relative risk. Relatively speaking, there was a more noticeable difference in point estimates under the proposed method compared to the existing method. For example, the reduction in RR under the joint intervention was estimated to be about 50\% higher using the proposed method compared to the existing method. However, the absolute values of these reductions were quite small at 0.038 and 0.025, and this difference is not clinically meaningful. These results suggest that when the effects of interventions are quite small and the mediators exhibit a weak correlation between the error terms in the joint mediator model (close to 0.3), there may not be measurable differences in the conclusions drawn using either method. This confirms what we observed in the simulation study for scenarios with small effects, low correlation, and no interactions.
\section{Discussion}
\label{section:discussion}
In this paper, we introduced a new method for performing a causal decomposition analysis in the presence of mediator variables that are correlated, contemporaneous in nature, and that may interact with one another to affect the outcome. This work extends the Monte Carlo-based causal decomposition analysis method by \cite{sudharsanan2021educational} by specifying a flexible joint mediator model and identifying the causal assumptions necessary to compute path-specific effects. Our method can be used with any combination of continuous and binary mediator variables, making it highly compatible with many types of data.

In the simulation study, our proposed method exhibited marked improvements over the existing method in terms of bias, average interval width, and coverage in the setting of two moderately correlated, continuous mediators. The performance was even stronger in the presence of a mediator-mediator interaction. There are many different settings were two moderately correlated (and interacting) continuous mediators might be of interest from studies of health behaviors \citep{schuit2002clustering, laaksonen2002associations} to studies of power plant emissions \citep{kim2019bayesian}. Another specific setting might include examining the mediating effect of diet scores and hours of physical activity per week. The data collected on physical activity levels in the REGARDS dataset were categorized as low, moderate, or high levels of physical activity and could be subject to recall bias due to the way in which it was collected. Therefore, we were not able to explore this relationship in our data analysis. However, this would be an interesting combination of continuous mediator variables to explore in a future decomposition analysis.

Our method almost always performed better than the existing method in terms of bias when either one binary and one continuous mediator was considered or when two binary mediators were considered; however, for certain effects, there was a smaller improvement in relative bias between the proposed and existing methods in these settings, or the existing method produced slightly less biased effects. The difference in average interval width was also less pronounced between the two methods in the mixed mediator or two binary mediator settings. Importantly though, across all scenarios, applying our proposed method in the absence of residual correlation had little effect on interval width and little effect on bias across all of the decomposition effect estimates. The downside of applying the proposed method when the existing method may be more appropriate is that the proposed method is more computationally intensive due to fitting and simulating from multivariate models.

The causal mediation and decomposition analysis literature largely focuses on assessing the effect of just one mediator variable or the joint effect of many mediator variables. However, there have been several recent papers focused on obtaining path-specific effects in mediation analyses \citep{wang2013estimation, lange2014assessing, kim2019bayesian, jerolon2020causal, oconnell2021pathway} and decomposition analyses \citep{sudharsanan2020rural}. Path-specific effects can more easily be tied to feasible interventions and can help determine which mediators (or combination of mediators) are contributing the most to explaining the difference in health outcomes between groups. To the best of our knowledge, there are no causal decomposition analysis methods that include multiple mediators while also accounting for the correlated nature of these mediators.

Our proposed methodology for causal decomposition analysis with multiple correlated mediators has several strengths. First, it builds upon the highly flexible simulation-based method proposed by \cite{sudharsanan2021educational}, and does not place requirements on the type of outcome model specified or the way in which relationships between the mediators and outcome are specified. Like existing causal decomposition analysis methods, it also focuses on computing observed differences, rather than causal differences, in the health outcomes by groups. It is highly applicable to studies of health disparities, which typically compare the outcomes between two groups that are defined based on an intrinsic, non-manipulable characteristic. The mediator-outcome relationships are the only causal relationships assumed, relaxing some of the strong no unmeasured confounding assumptions often required to conduct a mediation analysis. It can be easily implemented in the R software and is flexible in terms of the types of mediators included, while accounting for the residual correlation in the error terms. This allows researchers to compute not just joint, but also path-specific decomposition effects, if the causal identification assumptions are reasonable.

In addition to the strengths of the proposed causal decomposition analysis method, there are a number of limitations. First, this method may become computationally expensive or even infeasible to implement with too many mediator variables, since it involves estimating and simulating from a covariance matrix corresponding to all of the mediator variables' error terms. In the implementation of this approach in R that we provide in the supplementary materials, there are restrictions on the number of mediator variables that can be included. The \texttt{gjrm} function is currently written to fit joint mediator models with 2-3 continuous variables, 2-3 binary variables, or two variables where one is continuous and one is binary \citep{marra2017joint}. It is not possible to implement other combinations of mediator types at this time. Our method may be implemented in SAS by fitting the joint mediator model using \texttt{PROC QLIM}, which does not have restrictions on the number of mediator variables. However, we expect that it may not be feasible to implement this approach in SAS with more than four mediators for computational reasons. Finally, the causal assumptions must be carefully considered, particularly the assumptions of no unmeasured confounding between the mediators and outcome and the assumption that the covariance matrix does not change with a change in group label. The latter may be tested, but there is no clear rule as to what constitutes a ``similar'' covariance matrix between groups. In future work, we aim to explore how much of an impact differing covariance matrices have on the results of a decomposition analysis using our proposed method.

The causal decomposition analysis method we have presented in this paper can be used for planning interventions on modifiable mediator variables either individually or jointly. Many mediator variables are correlated in nature, especially those pertaining to lifestyle variables, environmental exposures, or social determinants of health. By testing potential interventions using our proposed causal decomposition analysis method, researchers and policymakers can form an evidence-based plan to intervene on the causes of differences in health outcomes between groups with the goal of reducing the gap in health outcomes between populations.

\section{Acknowledgements}
This research project is supported by cooperative agreement U01 NS041588 co-funded by the National Institute of Neurological Disorders and Stroke (NINDS) and the National Institute on Aging (NIA), National Institutes of Health, Department of Health and Human Service. The content is solely the responsibility of the authors and does not necessarily represent the official views of the NINDS or the NIA.  Representatives of the NINDS were involved in the review of the manuscript but were not directly involved in the collection, management, analysis or interpretation of the data.  The authors thank the other investigators, the staff, and the participants of the REGARDS study for their valuable contributions. A full list of participating REGARDS investigators and institutions can be found at:    \href{https://www.uab.edu/soph/regardsstudy/}{https://www.uab.edu/soph/regardsstudy/}


\newpage
\section*{Appendix}
\subsection*{Derivation of true effects in simulation study\label{app1}}
\begin{center}
\includegraphics[scale = 0.7]{simulation_dag.png}
\end{center}
\begin{align}
&E(Y(A, \boldsymbol{M} \sim f_{\boldsymbol{M}\mid a_1, a_2, c})) = \int E(y\mid a, m_1, m_2, c)dF(m_1, m_2, c \mid a_1, a_2)\\
= &\int E(y\mid a, m_1, m_2, c, u_2)dF(m_1, m_2, c, u_2 \mid a_1, a_2)\\
= &\int E(y\mid m_1, m_2, c, u_2)dF(m_1, m_2, c, u_2 \mid a_1, a_2)\\
= &\int E(y\mid m_1, m_2, c, u_2)dF(m_1, m_2, c, u_2, u_1 \mid a_1, a_2)\\
= &\int E(y\mid m_1, m_2, c, u_2)dF(m_1, m_2\mid c, u_1, a_1, a_2)dF(u_2\mid a)dF(c)dF(u_1)\\
= &\int E(y\mid m_1, m_2, c, u_2)dF(m_1 \mid c, u_1, a_1)dF(m_2 \mid c, u_1, a_2)dF(u_2\mid a)dF(c)dF(u_1)\\
= &\int E(y\mid m_1, m_2, c, u_2)dF(m_1 \mid c, u_1, a_1)dF(m_2 \mid c, u_1, a_2, l)dF(l\mid a_2)dF(u_2\mid a)dF(c)dF(u_1)
 \end{align}

\noindent Step (1): conditional exchangeability and counterfactual consistency assumptions. Step (2): Add in $U_2$ then integrate over it, since this still preserves the mediator-outcome causal relationships. Step (3): $Y$ is only dependent on $A$ through $U_2$, $M_1$, and $M_2$. Step (4): $Y \perp U_1 \mid \{M_1, M_2, C, U_2\}$, so we can integrate the expectation over $U_1$ as well. Step (5): Factor the joint distribution of $M_1$, $M_2$, $C$, $U_1$, and $U_2$ based on the relationships in the DAG. Step (6): Factor $f(m_1, m_2\mid c, u_1, a_1, a_2)$ into separate distributions conditional on the unmeasured confounder. Step (7): Add indirect pathway from $A$ to $M_2$ through $L$ then integrate out $L$.

To compute each true effect, we used Monte Carlo integration along with the selected simulation model parameters to evaluate the necessary integrals by setting $A$, $a_1$, and $a_2$ to the desired values. By allowing $a_1$ and $a_2$ to be any combination of 0 and 1, we implicitly assumed that the values of $a_1$ and $a_2$ do not affect $u_1$, which is causing the residual correlation between $M_1$ and $M_2$.
\subsection*{Additional figures\label{app2}}
\begin{figure*}[ht]
    \centering
    \includegraphics[scale = 0.1]{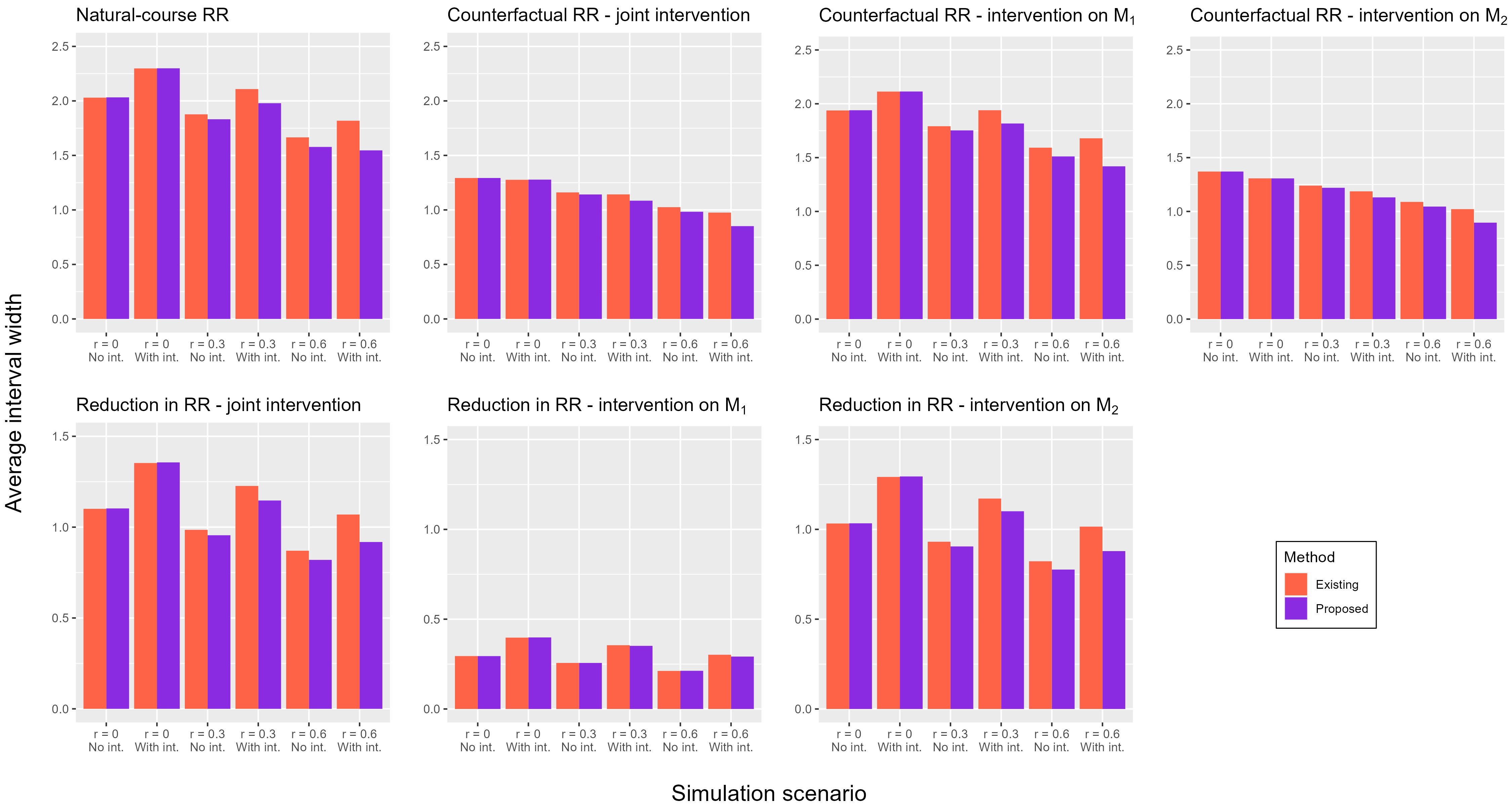}
    \caption{Average 95\% confidence interval width for each decomposition effect when $M_1$ was binary and $M_2$ was continuous. Relative risk was used to summarize the health differences between groups. $r$ is the approximate residual correlation and `int' indicates a mediator-mediator interaction effect.}
\end{figure*}
\begin{figure*}[ht]
    \centering
    \includegraphics[scale = 0.1]{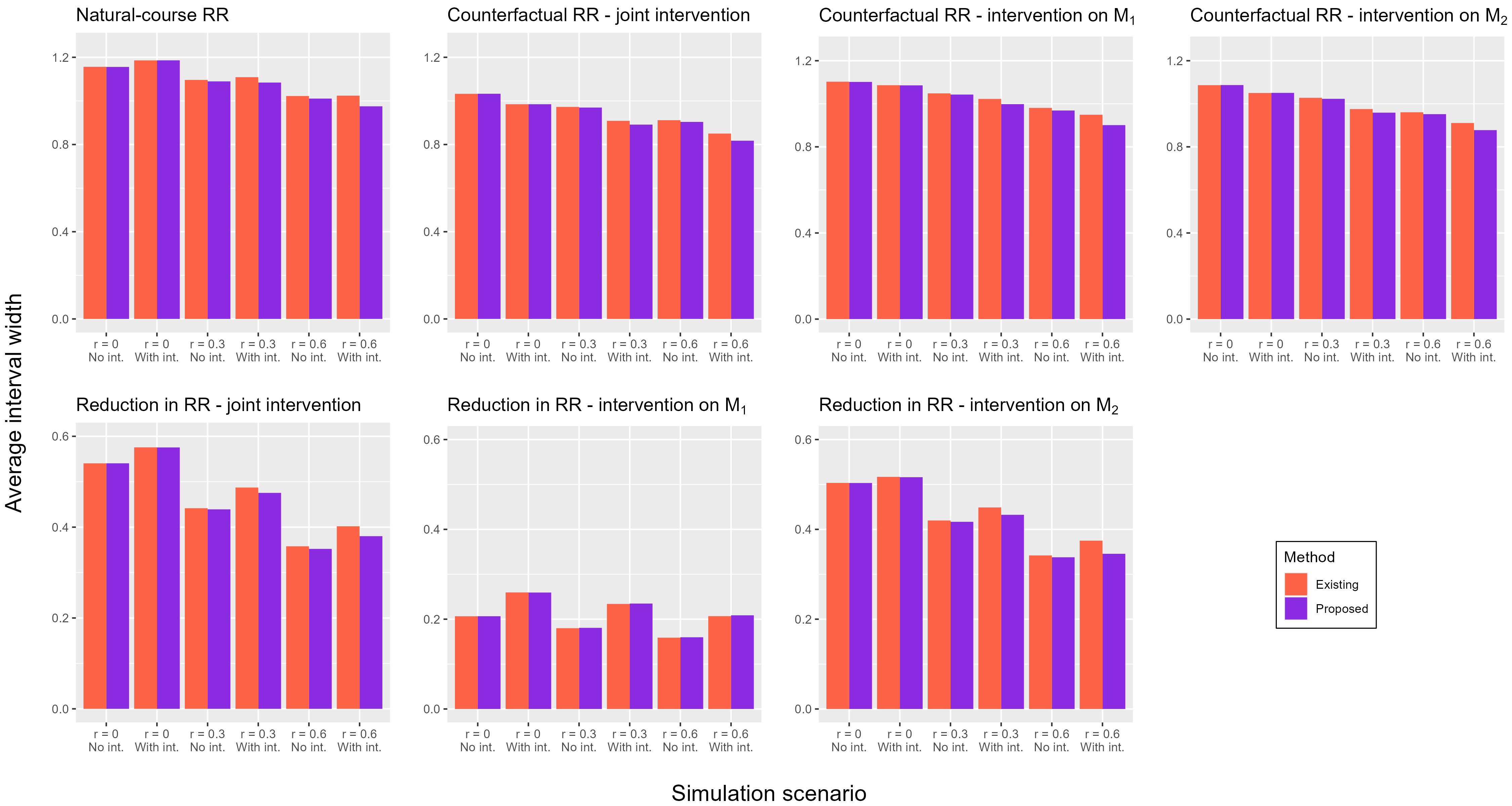}
    \caption{Average 95\% confidence interval width for each decomposition effect when both mediators were binary and the relative risk was used to summarize the health differences between groups. $r$ is the approximate residual correlation and `int' indicates a mediator-mediator interaction effect.}
\end{figure*}
\begin{figure*}[ht]
    \centering
    \includegraphics[scale = 0.1]{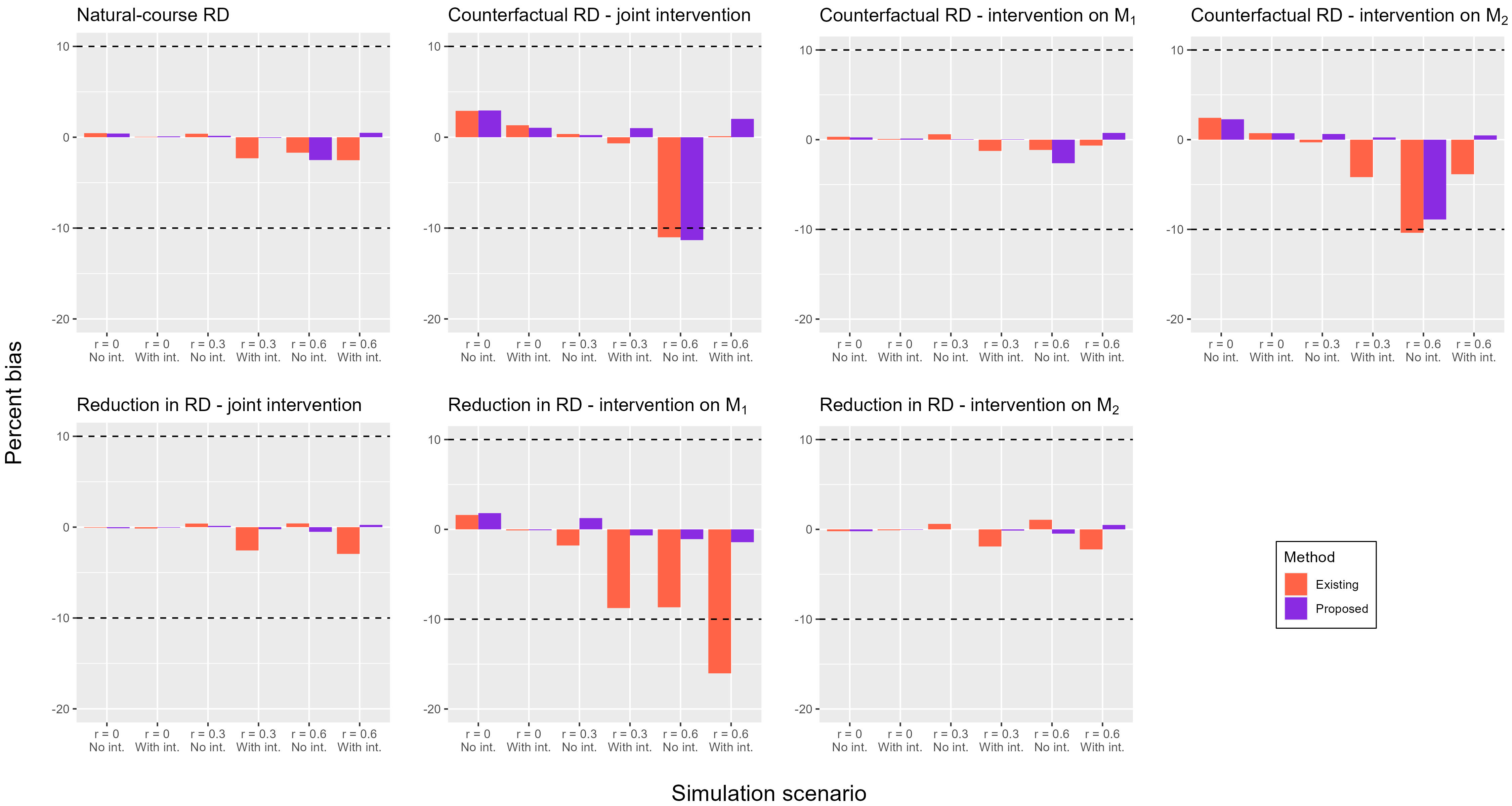}
    \caption{Percent bias for each decomposition effect when $M_1$ was binary and $M_2$ was continuous. Risk difference was used to summarize the health differences between groups. $r$ is the approximate residual correlation and `int' indicates a mediator-mediator interaction effect.}
\end{figure*}
\begin{figure*}[ht]
    \centering
    \includegraphics[scale = 0.1]{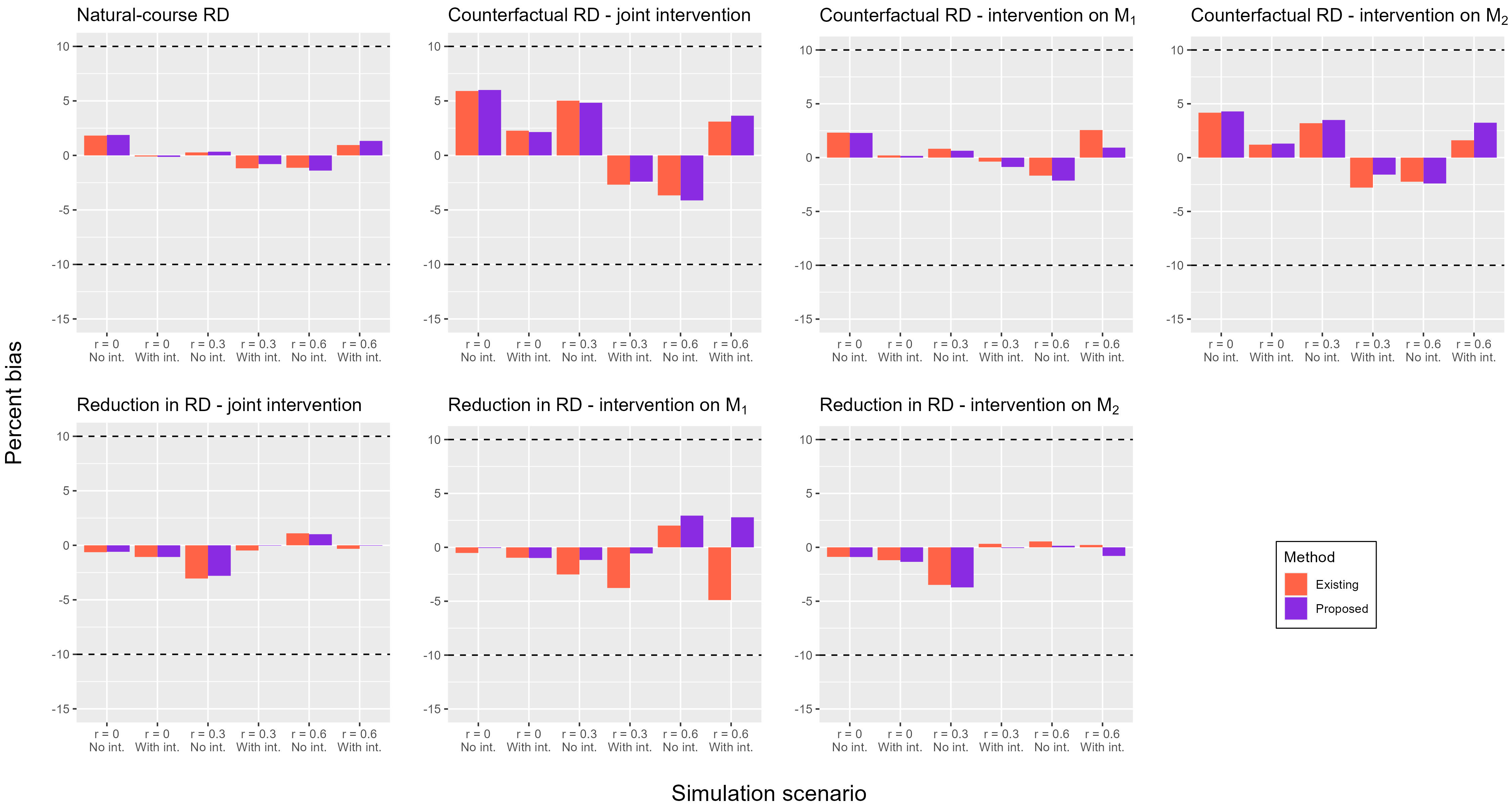}
    \caption{Percent bias for each decomposition effect when both mediators were binary and the risk difference was used to summarize the health differences between groups. $r$ is the approximate residual correlation and `int' indicates a mediator-mediator interaction effect.}
\end{figure*}

\end{document}